\documentclass{aa}

\usepackage{graphicx}

\usepackage{natbib}
\bibpunct{(}{)}{;}{a}{}{,}

\newcommand{\dd}{\, {\rm d}}	    
\newcommand{\env}{{\rm env}}	    
\newcommand{\disc}{{\rm disc}}	    
\newcommand{\rcyl}{r_{\rm cyl}}	    
\newcommand{\teff}{T_{\rm eff}}	    
\newcommand{\K}{\, {\rm K}}	    
\newcommand{\nm}{\, {\rm nm}}	    
\newcommand{\mm}{\, {\rm mm}}	    
\newcommand{\cm}{\, {\rm cm}}	    
\newcommand{\mum}{\, {\rm \mu m}}	    
\newcommand{\msun}{\, {\rm M}_\odot}	    
\newcommand{\lsun}{\, {\rm L}_\odot}	    
\newcommand{\rsun}{\, {\rm R}_\odot}	    
\newcommand{\au}{\, {\rm AU}}	    
\newcommand{\Jy}{\, {\rm Jy}}	    
\newcommand{\beam}{\, {\rm beam}}	    

\begin{document}

\title{The circumstellar disc in the Bok globule CB~26}
\subtitle{Multi-wavelength observations and modelling of the dust disc and envelope}
\author{
  J\"urgen Sauter \inst{1,2}
  \and Sebastian Wolf \inst{1}
  \and Ralf Launhardt \inst{2}
  \and Deborah L. Padgett \inst{3}
  \and Karl R. Stapelfeldt \inst{4}
  \and Christophe Pinte \inst{5,6}
  \and Gaspard Duch\^ene \inst{7,6}
  \and Fran\c cois M\'enard \inst{6}
  \and Caer-Eve~McCabe  \inst{3}
  \and Klaus Pontoppidan \inst{8}
  \and Michael Dunham \inst{9}
  \and Tyler L. Bourke \inst{10}
  \and Jo-Hsin Chen \inst{9}
}

\institute{
  Christian-Albrechts-Universit\"at zu Kiel, Institut f\"ur Theoretische Physik und Astrophysik, Leibnizstr. 15, 24098 Kiel, Germany
  \and Max-Planck-Institut f\"ur Astronomie, K\"onigstuhl 17, 69117 Heidelberg, Germany
  \and California Institute of Technology, 1200 E California Blvd, Mail code 220-6, Pasadena, CA 91125, USA
  \and JPL, 4800 Oak Grove Drive, Mail Stop 183-900, Pasadena, CA 91109, USA
  \and School of Physics, University of Exeter, Stocker Road, Exeter EX4 4QL, UK
  \and Laboratoire d'Astrophysique de Grenoble, CNRS/UJF UMR 5571, B.P. 53, F-38041 Grenoble Cedex 9, France
  \and Astronomy Department, University of California Berkeley, 601 Campbell Hall, Berkeley CA 94720-3411 USA 
  \and California Institute of Technology, Division of Geological and Planetary Sciences, MS 150-21, Pasadena, CA 91125, USA
  \and Department of Astronomy, University of Texas at Austin, 1 University Station C1400, Austin, TX 78712, USA
  \and Harvard-Smithonian Center for Astrophysics, 60 Garden St., Cambridge MA02138, USA
}

\date{Submitted: April 27, 2009 / Accepted July 6, 2009}

\abstract{
Circumstellar discs are expected to be the nursery of planets.
Grain growth within such discs is the first step in the planet formation process.
The Bok globule CB~26 harbours such a young disc.
} { 
We present a detailed model of the edge-on circumstellar disc and its envelope in the Bok globule CB~26.
} {
The model is based on HST near-infrared maps in the I, J, H, and K bands, OVRO and SMA radio maps at $1.1\mm$, $1.3\mm$ and $2.7\mm$, and the spectral energy distribution (SED) from $0.9\mum$ to $3\mm$.
New photometric and spectroscopic data from the Spitzer Space Telescope and the Caltech Submilimeter Observatory have been obtained and are part of our analysis.
Using the self-consistent radiative transfer code MC3D, the model we construct is able to discriminate parameter sets and dust properties of both its parts, namely envelope and disc.
} {
We find that the disc has an inner hole with a radius of $45\pm5 \au$.
Based on a dust model including silicate and graphite the maximum grain size needed to reproduce the spectral millimetre index is $2.5\mum$.
Features seen in the near-infrared images, dominated by scattered light, can be described as a result of a rotating envelope.
} {
Successful employment of ISM dust in both the disc and envelope hint that grain growth may not yet play a significant role for the appearance of this system.
A larger inner hole gives rise to the assumption that CB~26 is a circumbinary disc.
}


\maketitle

\section{Introduction}

CB\, 26 is a small cometary-shaped Bok globule located about $10^\circ$ north of the Taurus/Auriga dark cloud at  a distance of 140\,pc \citep{l01}.
An IRAS point source (IRAS 04559+5200) at its southwest rim suggests an embedded Class I young stellar object (YSO) \citep{s04} source. 
\cite{l97} found an unresolved 1.3mm continuum source associated with the IRAS source.

Interferometric observations by \cite{l01} showed that the major fraction of thermal dust emission at millimetre wavelengths has its origin in a young circumstellar disc with a diameter of about 400\,AU and a mass of about $0.1 \, \rm{M}_\odot$.
This disc is seen almost edge-on and the central star is not visible directly.
However, the spectral energy distribution suggests a Class\,I YSO with $L\ge 0.5 \, {\rm L}_\odot$ \citep{s04}.
From the $^{13}$CO line emission and the Keplerian rotation curve, \cite{l01} derive a central stellar mass of $M_{\ast} = 0.5\pm 0.1\,\rm{M}_\odot$.
Furthermore, \cite{l08} detected a jet-like molecular outflow emanating perpendicular to the plane of the disc.
This outflow seems to be co-rotating with the disc. 

We present new observations and a model for this source that accounts for spatially resolved data sets over more than 3 orders of magnitude as well as for the unresolved SED of the object. 

Our modelling is based on spatially resolved maps of CB~26 in the millimetre regime from the Sub-millimetre Array (SMA) and the Owens Valley Radio Observatory (OVRO), high resolution images in the I,J,H, and K bands obtained with the the Near Infrared Camera and Multi-Object Spectrometer (NICMOS) at the  Hubble Space Telescope (HST) and the Advanced Camera for Surveys (ACS), which is as well an instrument of the HST.
The photometric data for the SED upon our model is based are provided by the Multiband Imaging Photometer for Spitzer (MIPS) and the Infrared Array Camera (IRAC) aboard the Spitzer Space Telescope (SST) and millimetre photometry.
We also obtained a spectrum with the Infra-Red Spectrograph (IRS) aboard the SST. 

In this paper, we use all available NIR to mm continuum data on CB26 to develop a self-consistent model of the source.
This model consists of two parts:
First, an optically thick dust disc which accounts for the dark lane seen in images obtained with the HST of the object and the significantly elongated intensity profiles in the mm range.
The second component is an optically thin envelope that reproduces the scattered light nebulosity.

While the millimetre observations are sensitive only to radiation being emitted from dust in the dense region within the disc, the near-infrared images are dominated by scattered stellar light from dust in the circumstellar envelope and the disc's upper optically thin layers, often referred to as the  ``disc surface''.
These observations trace different physical processes in different regions of the circumstellar environment, but they are both strongly related to the dust properties in the system.

Thus, we are in the position not only to model observations on the common basis of one set of parameters, but we are also able to investigate whether the dust properties are different in the disc and the envelope.
This has been suggested by investigations of the dust evolution in circumstellar discs where dust grain growth alters the dust grain properties in the circumstellar disc quite considerably whilst it is of less importance in the low-density envelope.
Evidence for grain growth has been found, for instance, for the circumstellar discs IM Lupi \citep{p08}, GG Tau \citep{d04,c07}, HH~30 \citep{w04}, IRAS 04302+2247 \citep{w03a}, and VV Serpens \citep{a08}.

In order to compare our model with the available observations, we use the self-consistent radiative transfer code \texttt{MC3D} \citep{w99,w03c} in a parameter space study on the free parameters of the model.
We aim at finding the best-fit model, which we define to be the model that reproduces certain predefined features among the observational data (such as width of the dark lane; see below for further details) best.

\section{Observations and data reduction}
In the case of the disc in the Bok Globule CB~26, we are in the fortunate situation of having a large variety of observational data at hand.
This data includes not only the spectral energy distribution (SED) from $0.9 \mum$ to $2.7 \mm$ but also resolved maps in the near-infrared and in the millimetre regime.
We will now briefly discuss those observations and the data reduction in this section.

\subsection{HST Imaging}
The NICMOS and ACS data were taken by the GEODE\footnote{Group for Edge-On Disc Exploration} team.
An overview of the complete program, its objects, and its objectives can be found in Padgett et al. (2009, in  preparation).

\subsubsection{ACS}
CB~26 was observed with the Advanced Camera for Surveys Wide Field Channel on 2005 September 09.
Two 1250sec exposures were made in the F814W filter ($\lambda = 0.80\mum$, $\Delta\lambda = 0.15\mum$), corresponding to Johnson I band.
With a pixel scale of $0.05^{\prime\prime}$, the ACS provides a field of view of $201^{\prime\prime}\times100^{\prime\prime}$.
The images were reduced, combined to reject cosmic rays, and corrected for geometric distortion by the STScI pipeline.
Residual hot pixels and cosmic rays were manually removed by replacing the affected pixels by local median values. 

\subsubsection{NICMOS}
CB~26 was observed using NICMOS on 2005 September 15 with the F110W ($\lambda = 1.12\mum$, $\Delta\lambda = 0.16\mum$ ), F160W ($\lambda = 1.60\mum$, $\Delta\lambda = 0.12\mum$ ) and F205W ($\lambda = 2.06\mum$, $\Delta\lambda = 0.18\mum$) filters on the NIC2 array.
With a pixel scale of $0.075^{\prime\prime}$, the NIC2 array provides a field of view of $19.2^{\prime\prime}$ on a side.
All data were taken in a MULTIACCUM step=128 sequence, with total exposure times of 512secs for the F110W data and 384secs for the F160W and F205W data.
The observations were dithered back and forth by $0.75^{\prime\prime}$ for bad pixel removal.

All three of the NICMOS data sets were taken less than 1900 secs after emergence from the South Atlantic Anomaly (SAA) and therefore suffer from significant cosmic ray persistence.
The data were re-reduced using the most recent calibration files available.
The standard STScI CALNICA pipeline was used as a basis for the re-reduction, in addition to which, we employed both the {\it biaseq} and {\it pedsub} IRAF routines to remove the pedestal effects visible in each of the quadrants.
The {\it biaseq} task removes the time-dependent variations in the bias level for each quadrant, and the {\it pedsub} task then removes the fixed bias offset.
After the data have had the pedestal effect removed, we used the IDL {\it SAACLEAN} program to model and iteratively remove the persistent cosmic rays in the image.
After emerging from the SAA, NICMOS powers back up and takes 2 dark frames.
These dark frames provide the model for the cosmic ray persistence pattern; {\it SAACLEAN} takes this model and iteratively
removes it from the data until the noise in the background reaches a global minimum.
Further details on this procedure can be found in \cite{b03} (ISR 2003-010).
Following the {\it SAACLEAN} procedure, the data are run through a second round of pedestal subtraction, have any remaining bad pixels removed, and are then run through the CALNICB pipeline procedure to generate the final mosaics.

\subsection{Spitzer photometry and spectroscopy}
Mid- and far-infrared photometry for CB~26 were retrieved from the Spitzer Archive.
IRAC observations were carried out under GTO program 94 (PI: Charles Lawrence).
The data were taken 2004 February 11, AOR key 4916224, using the high dynamic range mode and 5 dithered exposures of 30sec each.
Aperture photometry was measured from the Spitzer pipeline ``post-BCD'' mosaics (version 11.0.2), using a standard 10 pixel radius aperture for the source and background annulus with 12-20 pixel radius.
MIPS observations were made under GTO program 53 (PI: George Rieke) on 2005 March 08, AOR key 12020480.
A medium scan map covering 0.2$^{\circ}\times0.5^{\circ}$ was performed, providing multiple dithered 4sec exposures and total integration times of 168, 84, and 16 sec at 24, 70, and 160 $\mum$, respectively.
Spitzer post-BCD pipeline mosaics (version 14.4.0) were used for aperture photometry.
Photometry was measured in apertures whose radii and background annuli were 7$^{\prime\prime}/7^{\prime\prime}-13^{\prime\prime}$ (24$\mu$m), 35$^{\prime\prime}/39^{\prime\prime}-65^{\prime\prime}$ (70$\mu$m), and 48$^{\prime\prime}/64^{\prime\prime}-128^{\prime\prime}$ (160\,$\mu$m) with aperture correction factors as given by \cite{e07,g07,s07}.

We carried out our own Spitzer low-resolution spectroscopy of CB~26 on 2006 October 18 under program 30765, AOR key 18964992.
The ramp duration and number of observing cycles used were 6$\times$14 sec, 4$\times$14 sec, 8$\times$30 sec, and 8$\times$30 sec for the SL2, SL1, LL2, and LL1 modules of the IRS (respectively).
The spectra were processed beginning with the intermediate {\it droopres} products from pipeline version 15.3.0.
The 2D images were co-added and the nod pairs subtracted to remove stray light and other additive artifacts, as well as rogue pixels.
1D spectra of each nodding beam were then extracted using a fixed aperture of 5 pixels.
Following \cite{b08} the spectra were flux calibrated using a spectral response function determined by comparing standard star observations with template spectra.
These response functions are also used with the ``C2D'' Legacy data set \citep[e.g.][]{k06}.
It was not necessary to apply any scaling to match the short-low and long-low modules, indicating that the source is unresolved at the IRS wavelengths.

\subsection{Millimetre and Sub-millimetre measurements}

\subsubsection{SMA}

Observations at 270\,GHz ($1.1\mm$) with the Sub-Millimetre Array \citep[SMA,][]{h04} were made in December 2006, in two configurations providing baselines in the range 12\,--\,62\,k$\lambda$.
Typical system temperatures were 350--500\,K.
The quasar 3C279 was used for bandpass calibration, and the quasars B0355+508 and 3C111 for gain calibration.
Uranus was used for absolute flux calibration, which is accurate to 20-30\%.
The data were calibrated 
using the IDL MIR package (Qi 2005) and imaged using MIRIRAD \cite{s95}.  
The cleaned and restored 1.1\,mm continuum map was constructed with robust $uv$-weighting using line-free channels in both sidebands. 
Here we only use the continuum map.
The observations together with the molecular line data are described in more detail in a forthcoming paper (Launhardt et al. in prep.).

\subsubsection{OVRO}
CB\,26 was also observed with the Owens Valley Radio Observatory (OVRO) between January 2000 and December 2001. 
Four configurations of the six 10.4\,m antennas provided baselines in the range 6\,--\,180\,k$\lambda$\ at 2.7\,mm (110\,GHz) and 12\,--\,400\,k$\lambda$\ at 1.3\,mm (232\,GHz).
Average SSB system temperatures of the SIS receivers were 300\,--\,400\,K at 110\,GHz and 300\,--\,600\,K at 232\,GHz.
The raw data were calibrated and edited using the MMA software package \citep{s93}.
Mapping and data analysis used the MIRIAD toolbox.
Observing parameters are described in detail in \cite{l01}. 
The data presented here include additional observations conducted in 2001. 
All maps are generated with robust $uv$-weighting, cleared, and restored with a clean beam.
Effective synthesised beam sizes of all interferometric millimetre continuum 
maps used here are summarised in Table \ref{beamTable}.

\subsubsection{CSO}
Submillimeter observations of CB26 at 350 $\mu$m were obtained with the Submillimeter High Angular Resolution Camera II (SHARC-II) at the Caltech Submillimeter Observatory (CSO) on 2007 October 21.
SHARC-II is a $12 \times 32$ element bolometer array giving a $2.59\arcmin \times 0.97 \arcmin$ field of view \citep{d03}.
The beam-size at 350 $\mu$m is 8.5\arcsec.  

We used the Lissajous observing mode to map a region approximately 1\arcmin\ $\times$ 0.5\arcmin, centred at the position of the source.  
We obtained two scans, each 10 minutes long, for a total integration time of 20 minutes in good weather ($\tau_{225 \rm GHz} \sim 0.06$).
During both scans the Dish Surface Optimisation System (DSOS)\footnote{See \texttt{www.cso.caltech.edu/dsos/DSOS\_MLeong.html}} was used to correct the dish surface for gravitational deformations as the dish moves in elevation.

The raw scans were reduced with version 1.61 of the Comprehensive Reduction Utility for SHARC-II (CRUSH), a publicly available,\footnote{See \texttt{www.submm.caltech.edu/\~{}sharc/crush/index.htm}} Java-based software package.
CRUSH iteratively solves a series of models that attempt to reproduce the observations, taking into account both instrumental and atmospheric effects \citep{k06b,k06c,b06}.
Pointing corrections to each scan were applied in reduction based on a publicly available\footnote{See \texttt{www.submm.caltech.edu/\~{}sharc/analysis/pmodel/}} model fit to all available pointing data.
Pixels at the edges of the map with a total integration time less than 25\% of the maximum were removed to compensate for the increased noise in these pixels.
We then used Starlink's \emph{stats} package to assess the rms noise of the map, calculated using all pixels in the off-source regions.
The final map has a 1$\sigma$ rms noise of 70 mJy beam$^{-1}$.

Photometry was measured in a 20\arcsec\ aperture centred at the peak position of the source.
Calibration was performed according to the method used by \cite{s00,w07}.
This method is based on the  requirement that a point source should have the same flux density in all apertures with diameters greater than the beam FWHM (8.5\arcsec\ for these observations).
To briefly summarise, a flux conversion factor (FCF) is calculated for a 20\arcsec\ aperture by dividing the total flux density of a calibration source in Jy by the calculated flux density in the native instrument units of $\mu$V in a 20\arcsec\ aperture.
Flux densities of science targets are then derived by multiplying the 20\arcsec\ aperture flux density (in the instrument units) of the source by the FCF.
We measure a 350 $\mu$m flux density for CB~26 of 2.8 $\pm$ 0.6 Jy.

\section{Results from Observations - Basis for modelling}
The focus of this section are the results by the previously described observations.
The presented data set forms the basis of our modelling.

\subsection{Images}
\begin{figure}
  \resizebox{\hsize}{!}{
   \includegraphics[width=0.5\textwidth]{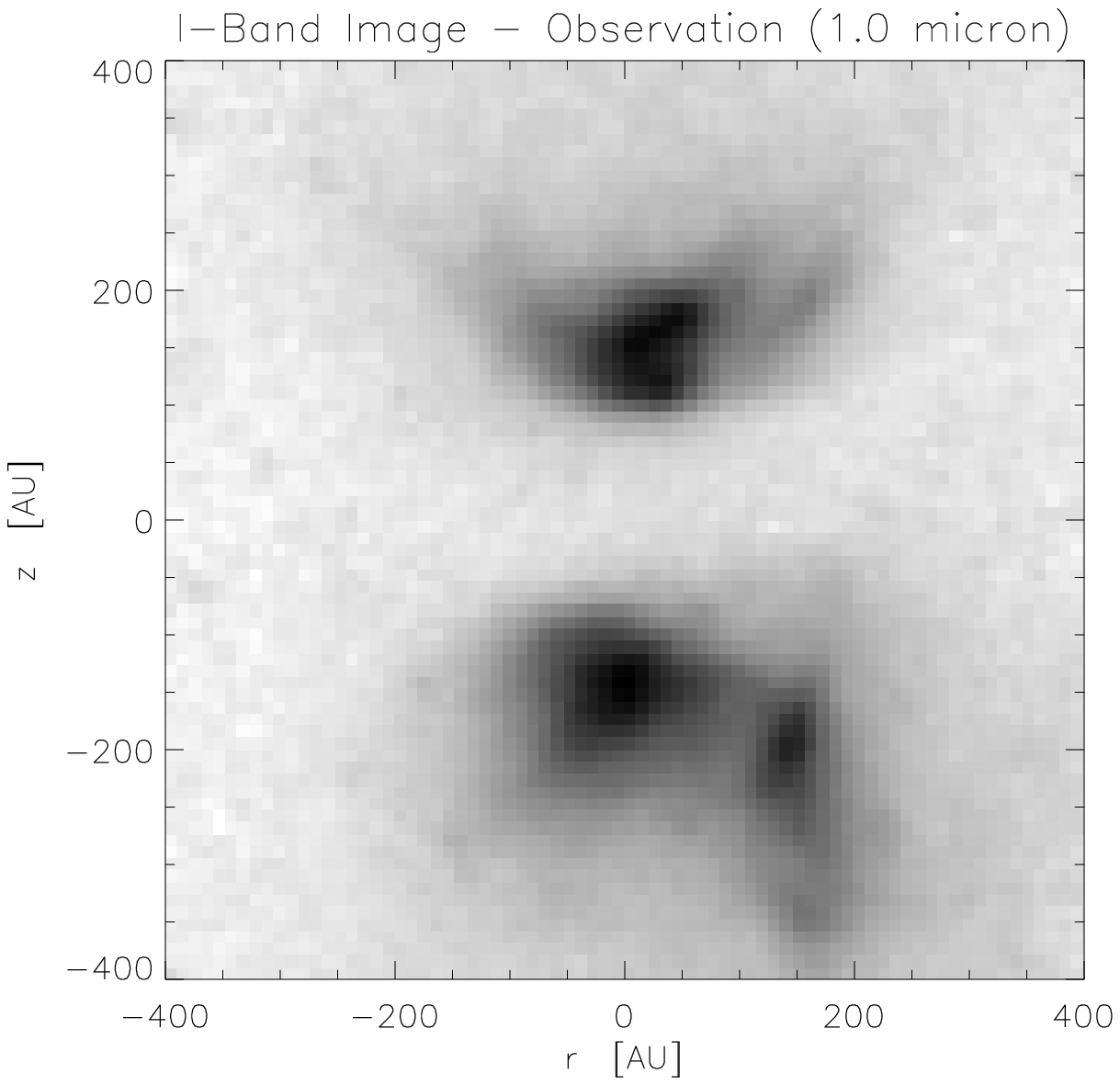}
   \includegraphics[width=0.5\textwidth]{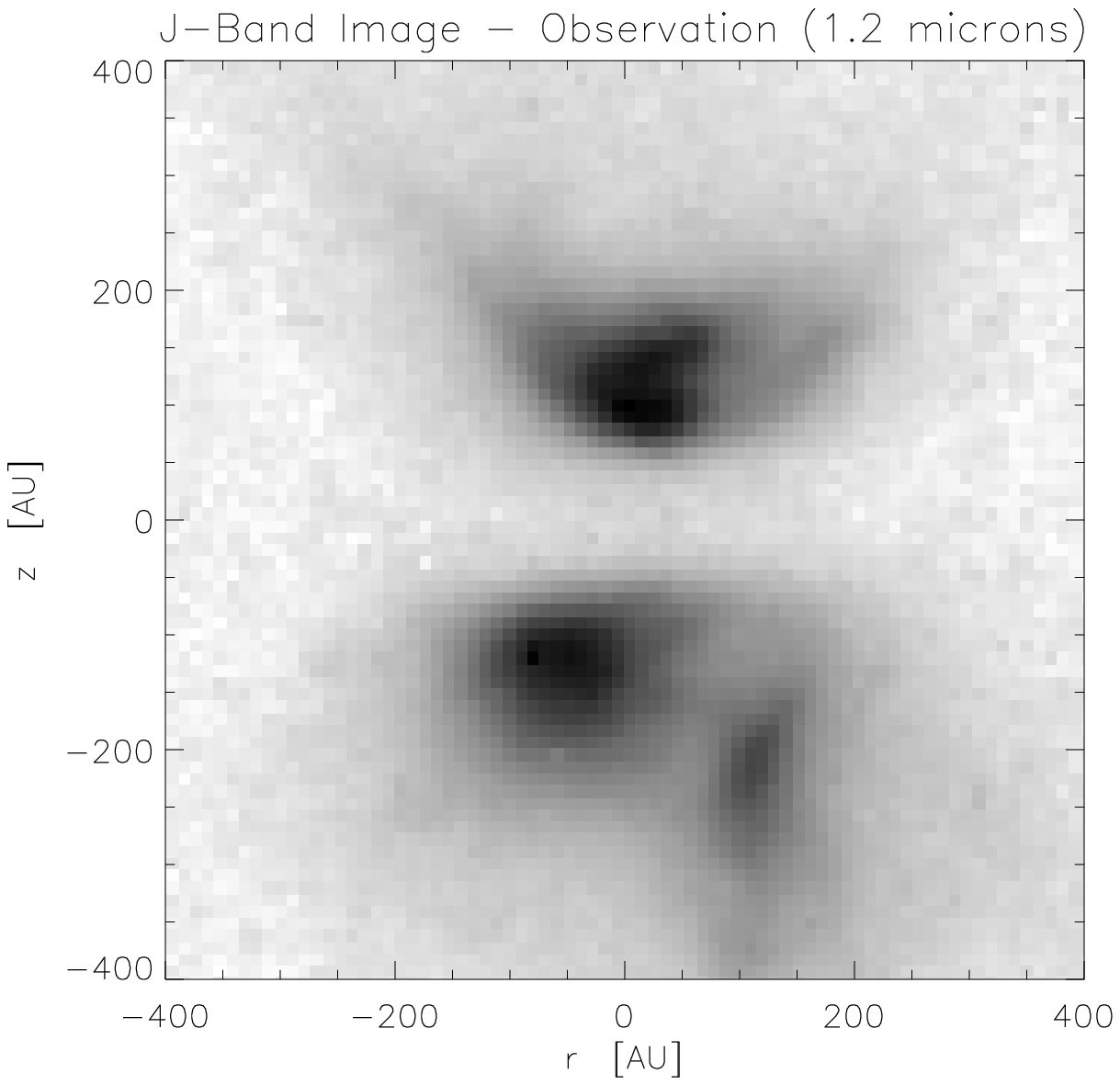}
  }
  \resizebox{\hsize}{!}{
   \includegraphics[width=0.5\textwidth]{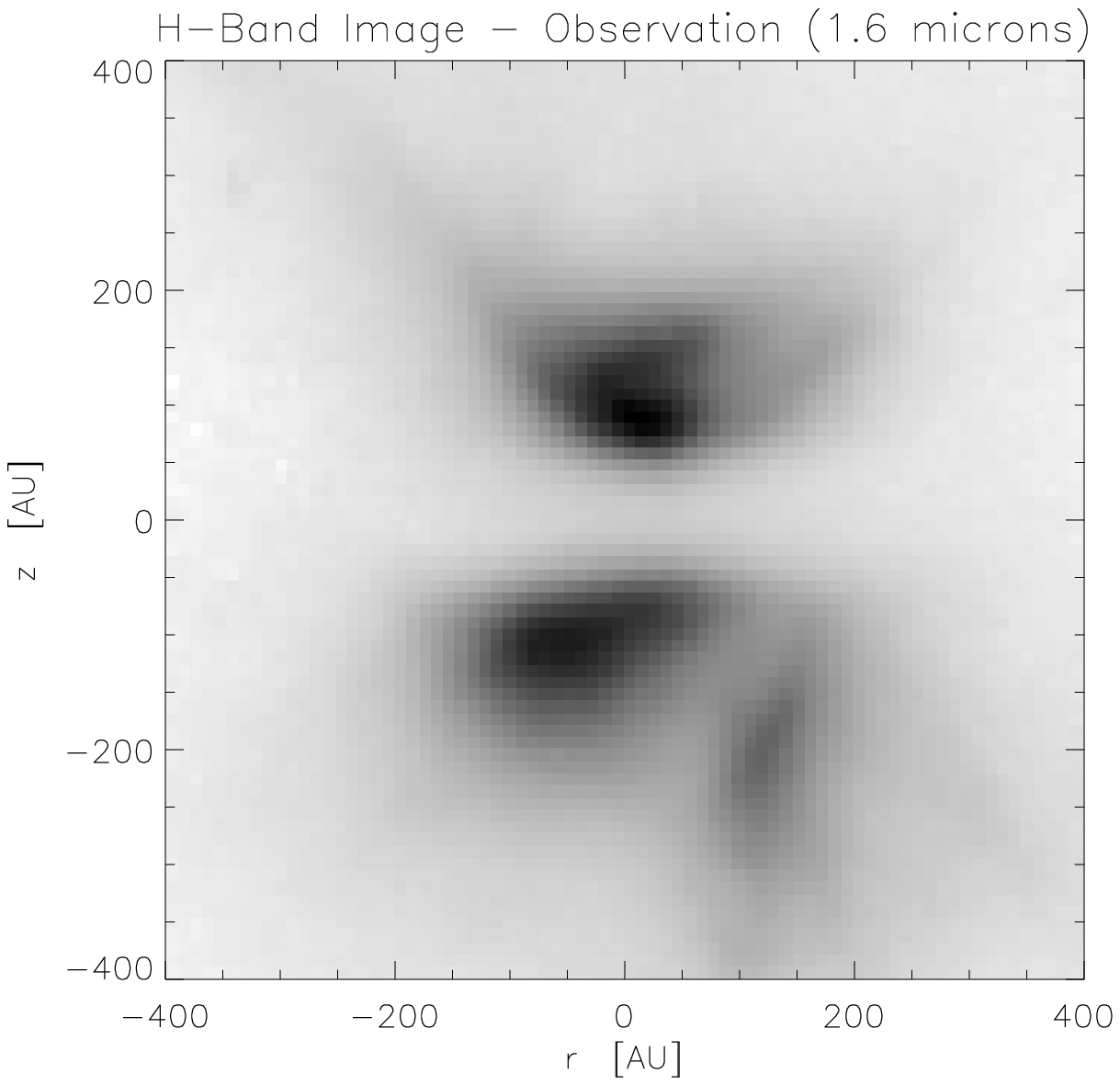}
   \includegraphics[width=0.5\textwidth]{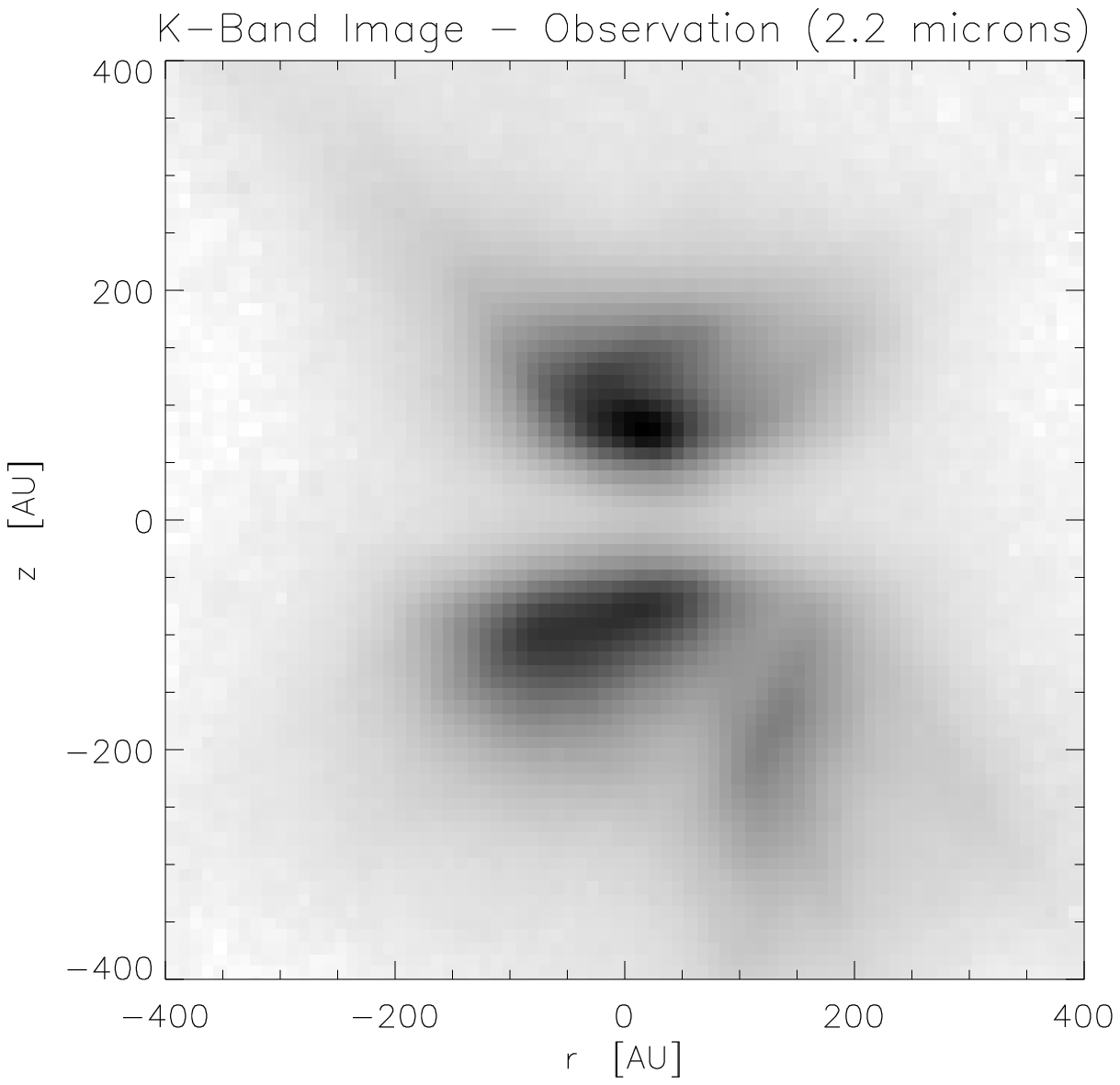}
  }
  \caption{
    Inverse HST images obtained with ACS and NICMOS in the I, J, H and K band.
    The colour scale is $\sim S^{\frac{2}{5}}$
    All four images have been rotated by $-30^\circ$ in order to align the major axis of the dark lane with the horizontal axis.
    For the image scale a distance to the object of 140\,pc is assumed ($100\au = 0.7^{\prime\prime}$).
    The image orientation is PA $330^\circ$ up, and PA $60^\circ$ to the left.
  }
  \label{NIRmaps}
\end{figure}

The maps from the HST in I, J, H, and K bands are shown in Fig. \ref{NIRmaps}.
All all four images have been rotated in order to align the major axis of the dark lane with the horizontal axis.
The emission seen on these maps is pure scattered light from the central star.
The dark shadowy line that intersects the bipolar structure is present in all images.
Especially the dependency of its width on the wavelength is clearly visible.
Since one expects the circumstellar disc to be optically thick at these wavelengths, the dark lane is interpreted as the disc's shadow in the encompassing envelope structure.
The bipolar nebula also shows a complex morphology far above and especially below the disc.
For further discussion we refer the interested reader to Padgett et al. (2009, in preparation).

\begin{figure}
  \resizebox{\hsize}{!}{
   \includegraphics[width=0.5\textwidth]{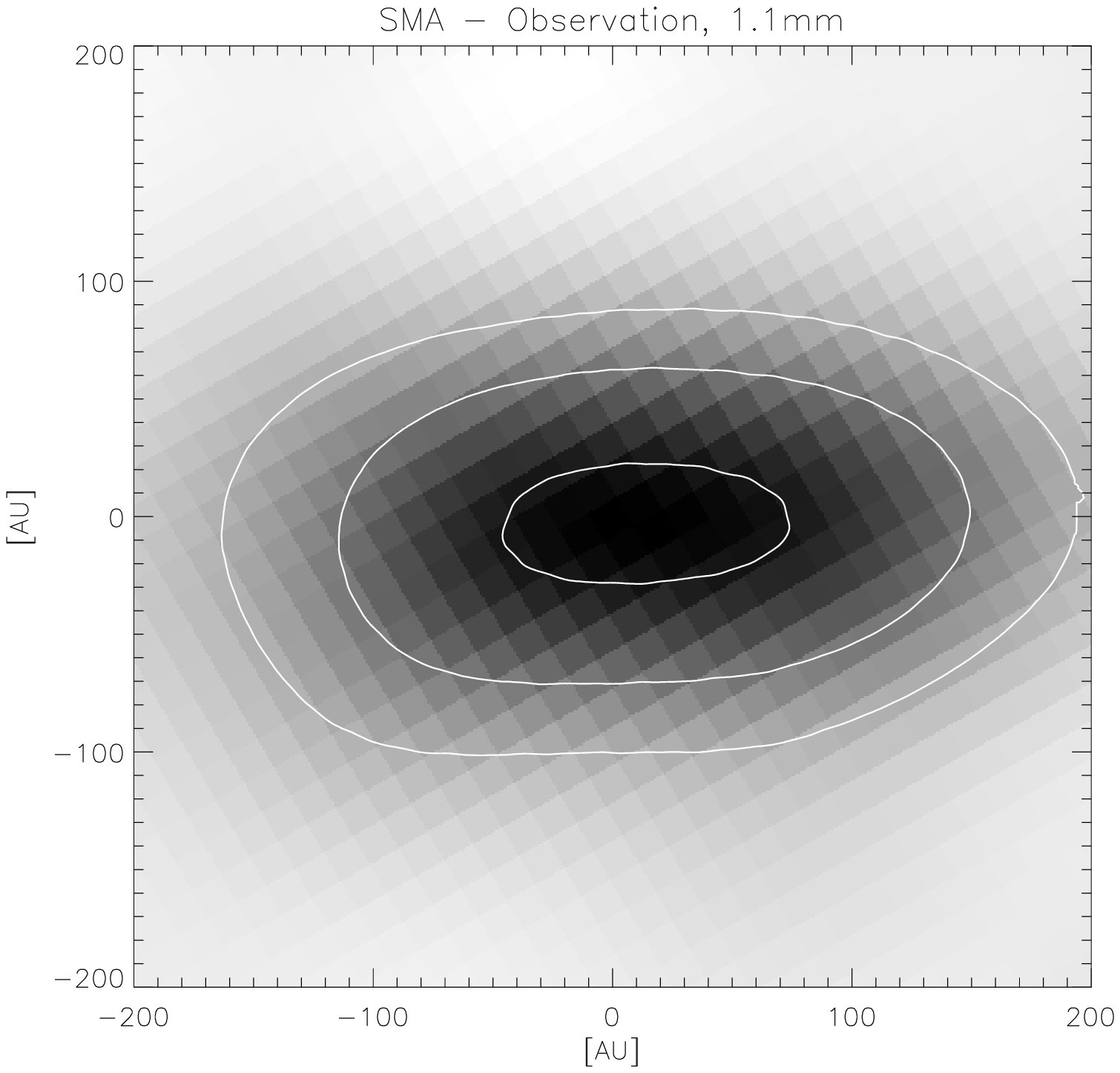}
   \includegraphics[width=0.5\textwidth]{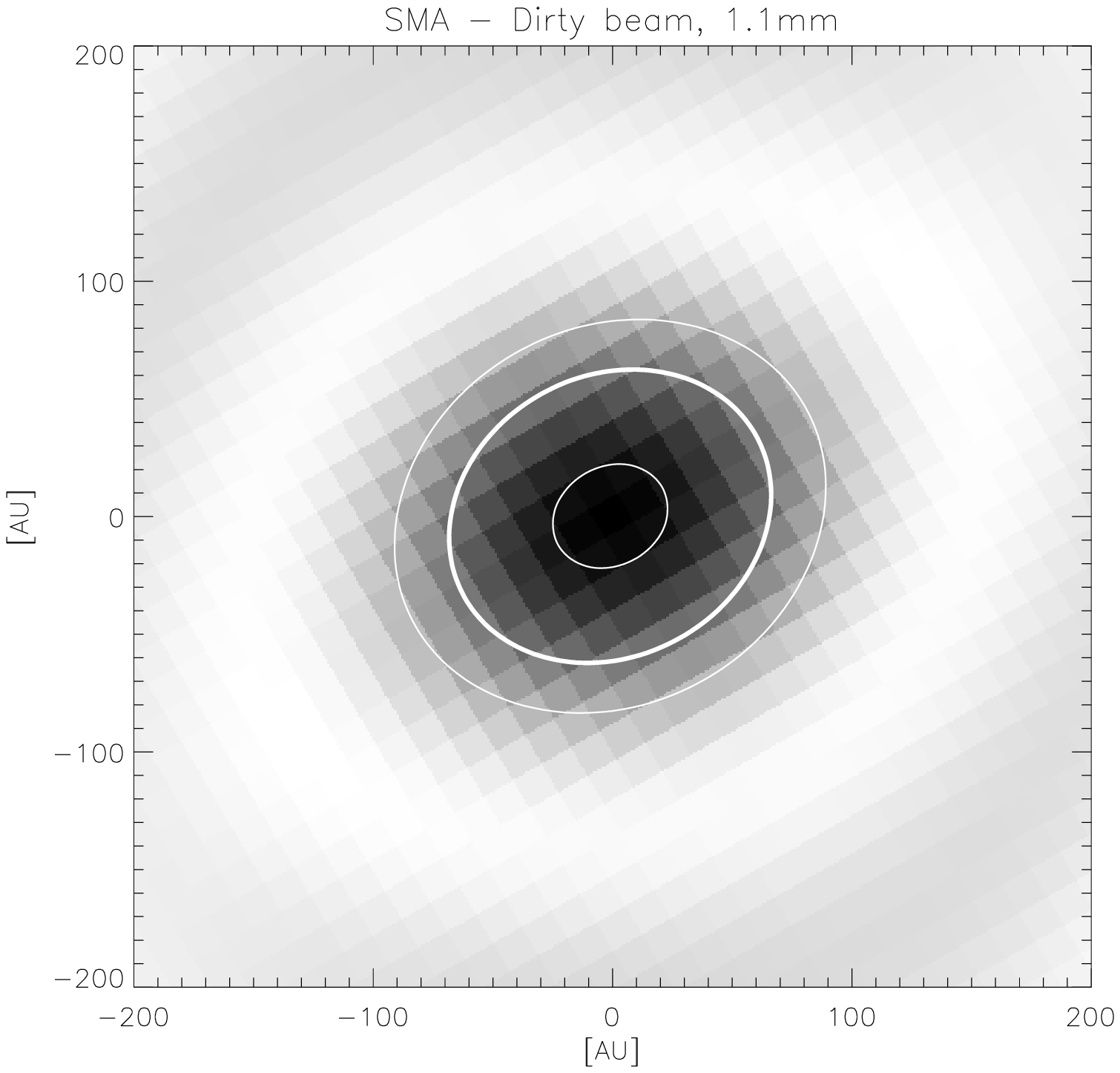}
  }
  
  \resizebox{\hsize}{!}{
   \includegraphics[width=0.5\textwidth]{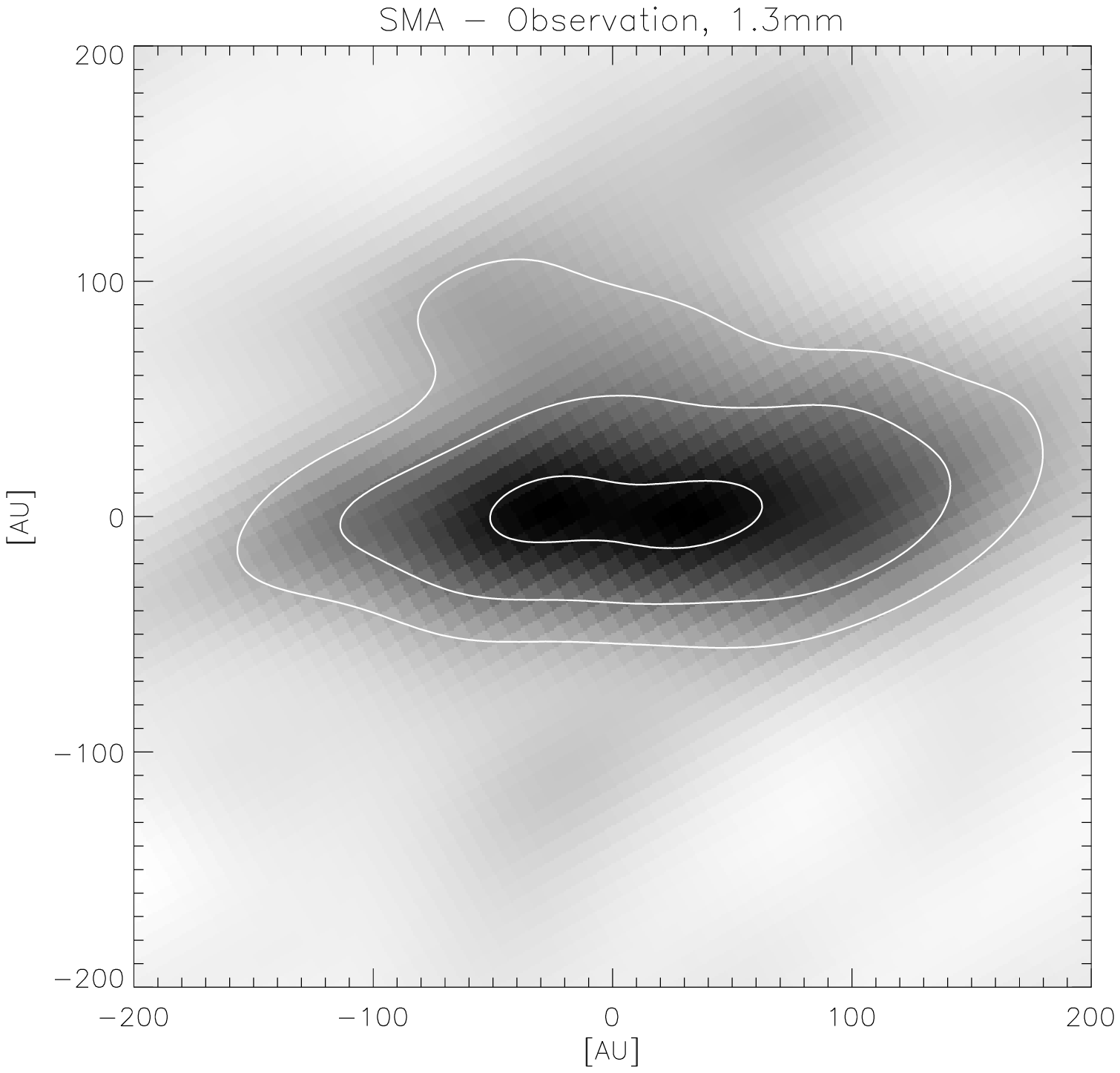} 
   \includegraphics[width=0.5\textwidth]{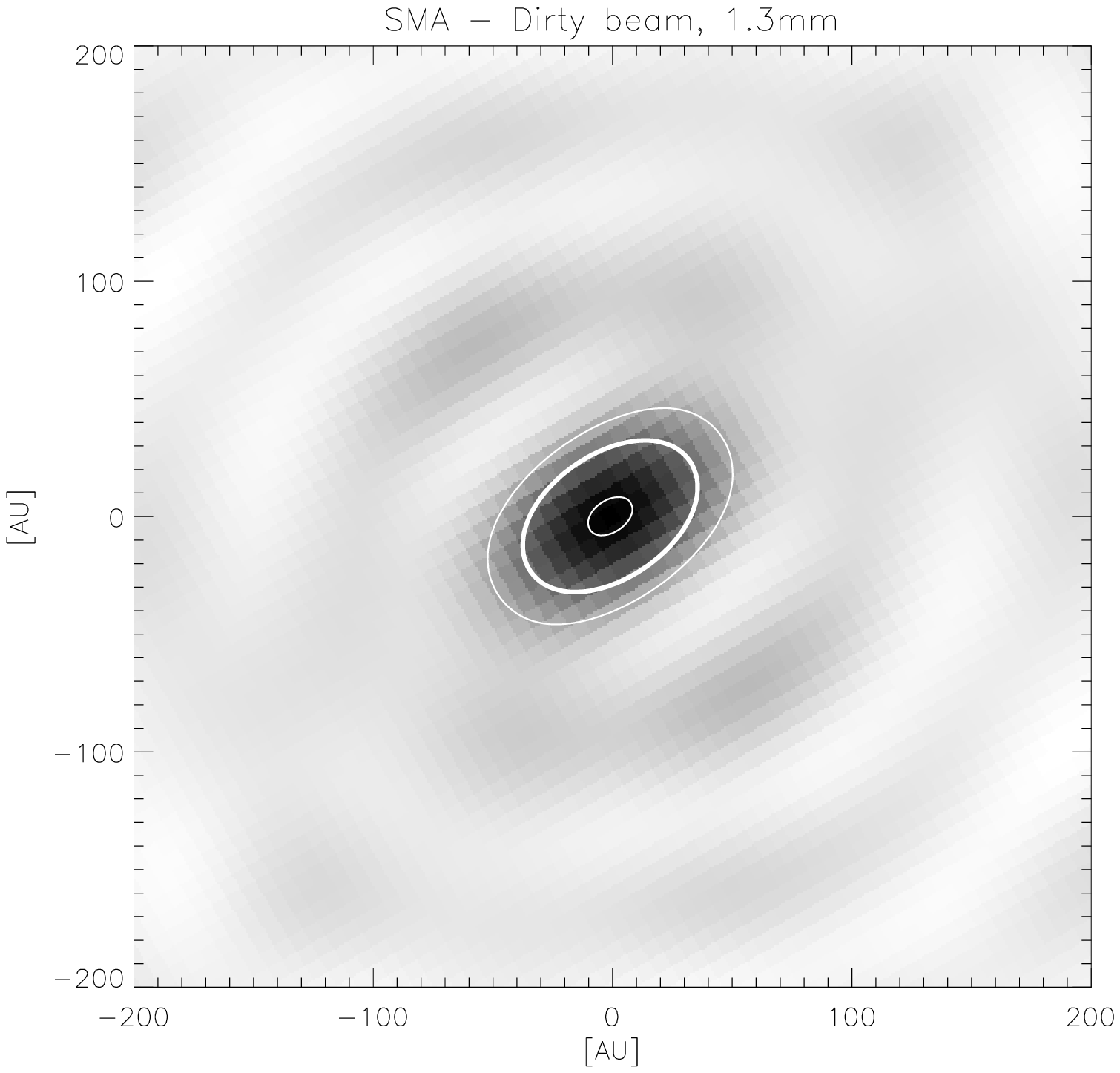}
   }
  \resizebox{\hsize}{!}{
   \includegraphics[width=0.5\textwidth]{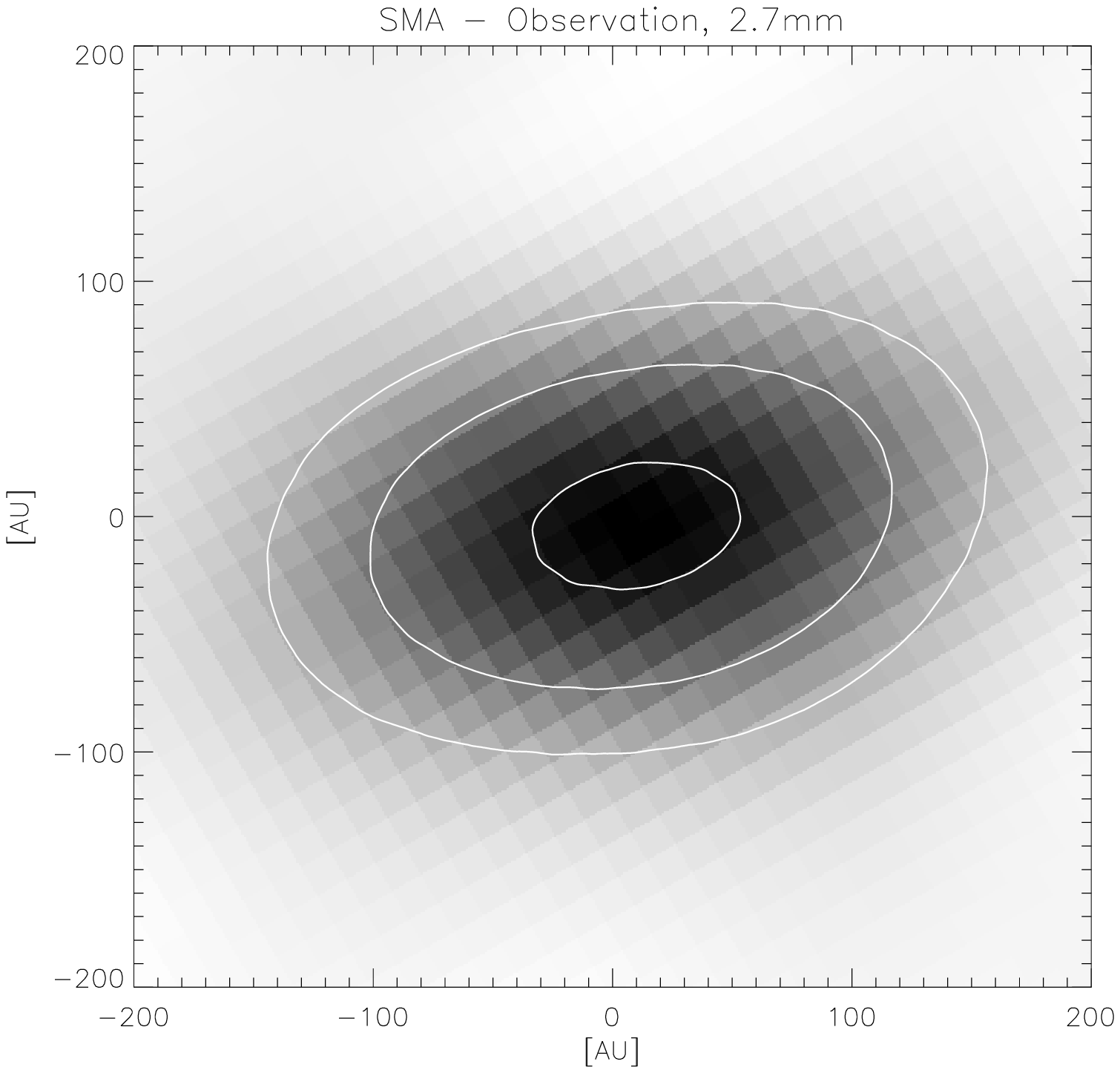} 
   \includegraphics[width=0.5\textwidth]{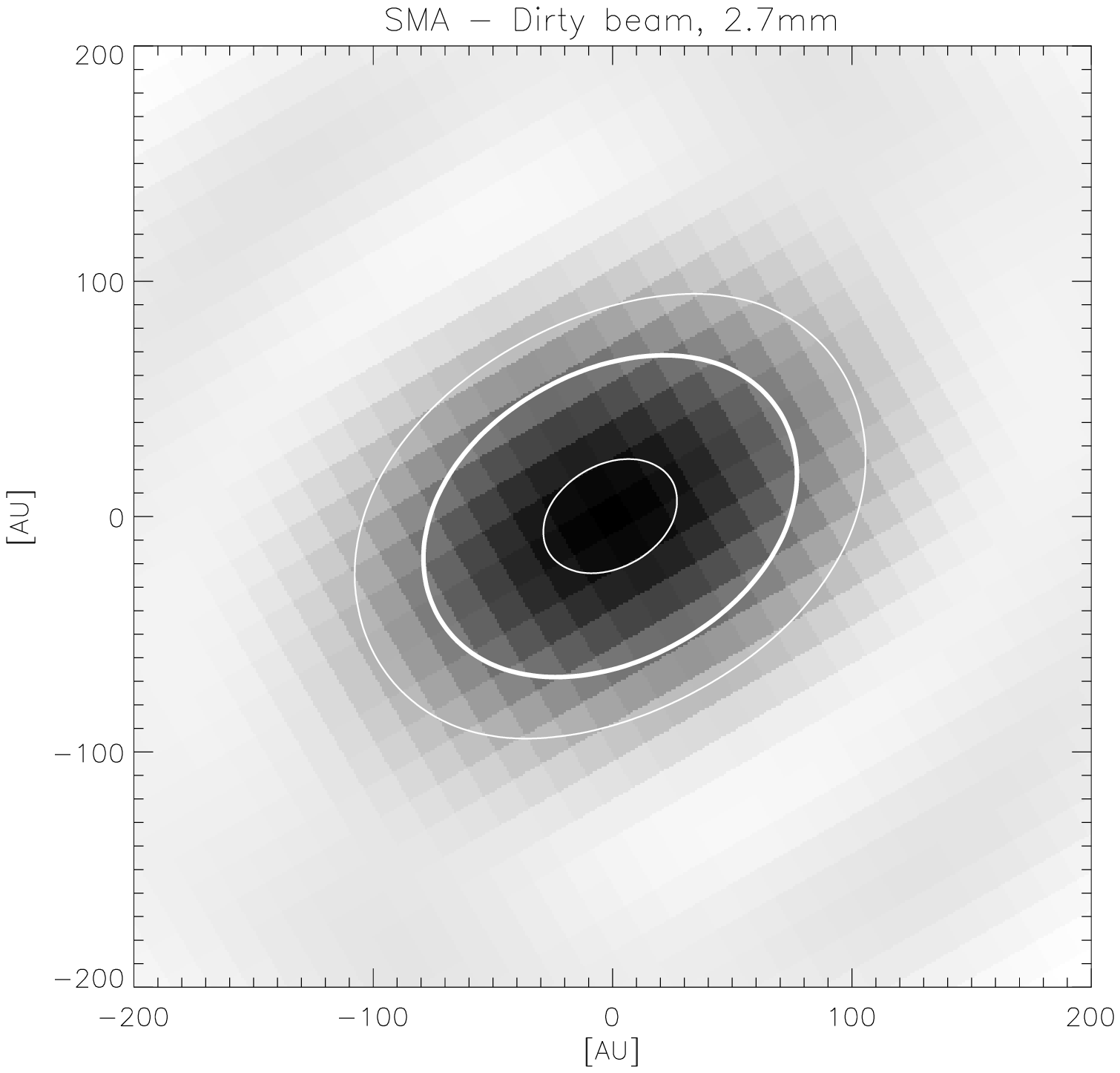} 
  }
  \caption{
    Inverse, reconstructed images from millimetre interferometric observations (\emph{left column}) and corresponding dirty beam maps (\emph{right column}),    linear colour scale.
    All images have been rotated by $-30^\circ$ in order to align the major axis of the brightness distribution with the horizontal axis.
    For the image scale a distance to the object of 140\,pc is assumed ($100\au = 0.7^{\prime\prime}$).
    The contour lines are drawn at (from inside out) 90\%, 50\%, and 25\% of the image maximum flux value.
    In the dirty beam images, the 50\% contour, marking the FWHM of the Gaussian clean beam, is marked bold.
  }
  \label{mmmaps}
\end{figure}

The interferometric millimetre continuum maps at 1.1\,mm, 1.3\,mm, and 2.7\,mm, together with the dirty beam maps (see also Table \ref{beamTable}), are shown in Fig. \ref{mmmaps}. 
For the ease of modelling, these images have been rotated $30^\circ$ in order to align the major axis of the elongated structure with the horizontal axis.
While the source is unresolved vertically in all images, it is well-resolved along its horizontal axis, especially in the highest-resolution map at 1.3\,mm. 
The 1.3\,mm map also recovers some extended emission from the envelope that might be related to a disc wind \citep[see][]{l01}.
For all images shown, radio and NIR, North is the same direction.

\begin{figure}
  \resizebox{\hsize}{!}{
    \includegraphics{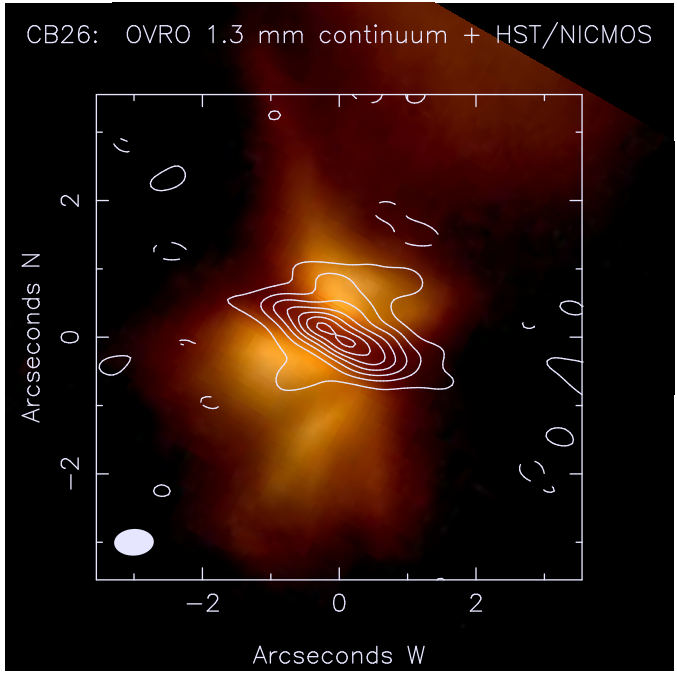}
  }
  \caption{Overlay of the OVRO $1.3 \mm$ continuum map and the HST NICMOS images.
	   The NICMOS colour image is a 3-colour composite of F205W, F160 W, and F110W (in the RGB colour planes, respectively), shown in log stretch.
	   The contour levels are linear, with the lowest contour at about the 2.5 $\sigma$-level, the others at the 6, 10, 14, 22, and 25 $\sigma$-level, respectively.
	  }
  \label{HSTradiooverlap}
\end{figure}

Fig. \ref{HSTradiooverlap} shows an overlay of a NIR colour-composite image and the $1.3\mm$ dust continuum emission. 
The spatial co-location of the millimetre dust emission and the dark lane in the scattered light images confirms the hypothesis of an edge-on optically thick disc as explanation for the observed features. 
However, the HST pointing is only good to about 1 arcsec. 
Due to the high extinction in the cloud core and the small NICMOS field of view, there are no reference stars in the image that could be used to correct for this pointing uncertainty. 

\begin{table}
  \label{beamTable}
  \caption{Overview of the beam sizes in the millimetre-maps}
  \centering
  \begin{tabular}{c c c c}
    \hline \hline
      Instrument & $\lambda [{\rm mm}]$ & PSF (FWHM) [$^{\prime\prime}$] & Orientation\\
    \hline
      SMA & $1.1$ & $1.00\times0.84$ & $-59.3^\circ$\\
      OVRO & $1.3$ & $0.61\times0.36$ & $-53.8^\circ$\\
      OVRO & $2.7$ & $1.21\times0.87$ & $-58.3^\circ$\\
    \hline
  \end{tabular}
\end{table}

\subsection{Spectral energy distribution}
\begin{figure}
  \resizebox{\hsize}{!}{
    \includegraphics{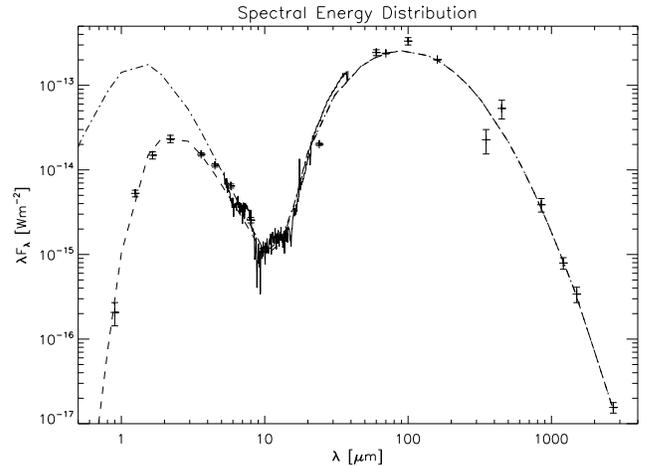}
  }
  \caption{Spectral energy distribution. Data with error-bars from Table \ref{sedTable}.
  The IRS spectrum is the solid line.
  The dashed-dotted line corresponds to the best-fit model.
  The dashed line is the best-fit model with the dust screen.}
  \label{SED}
\end{figure}

The results of photometric measurements are presented in Table \ref{sedTable} and Fig.\ref{SED}.
The spectrum we obtained from the IRS very well complements the photometric points as seen in Fig. \ref{SED}.

As is evident in this Figure, the 350 $\mu$m SHARC-II flux density is lower than expected based on comparison to the complete SED.
This discrepancy can be explained by the fact that the Lissajous observing mode is insensitive to extended emission, as noted by \cite{w07}.
This issue will be explored in more detail by M. Dunham et al. (2009, in preparation), but preliminary results suggest that flux densities measured from this particular observing mode may underestimate the true flux density by up to a factor of 2, even for relatively compact objects.

\begin{table*}
  \label{sedTable}
  \centering
  \begin{tabular}{c c c c c}
    \hline \hline
      $\lambda [\mum]$ & Flux [mJy] & Aperture [$ ^{\prime\prime}$] & Instrument & Reference \\
    \hline 
	$0.90$ &  $0.062  \pm 0.019 $ & $24  $ & CAHA 3.5m     & (1) \\
	$1.25$ &  $2.2    \pm 0.2   $ & $12  $ & CAHA 3.5m     & (1) \\ 
        $1.65$ &  $8.2    \pm 0.8   $ & $12  $ & CAHA 3.5m     & (1) \\ 
        $2.20$ &  $17.1   \pm 1.7   $ & $12  $ & CAHA 3.5m     & (1) \\
	$3.6 $ &  $18.3   \pm 0.8   $ & $12  $ & Spitzer IRAC1 &     \\
	$4.5 $ &  $17.0   \pm 0.8   $ & $12  $ & Spitzer IRAC2 &     \\
        $5.8 $ &  $12.5   \pm 0.7   $ & $12  $ & Spitzer IRAC3 &     \\
        $8.0 $ &  $ 6.8   \pm 0.5   $ & $12  $ & Spitzer IRAC4 &     \\
	$24  $ &  $160.6  \pm 5.2   $ & $30  $ & Spitzer MIPS1 &     \\
        $60  $ &  $4880   \pm 390.4 $ & $75  $ & IRAS PSC      &     \\
	$70  $ &  $5555   \pm 52    $ & $120 $ & Spitzer MIPS2 &     \\
	$100 $ &  $11100  \pm 1110  $ & $125 $ & IRAS PSC      &     \\ 
	$160 $ &  $10731  \pm 69.49 $ & $24  $ & Spitzer MIPS3 &     \\
	$350 $ &  $2650   \pm 850   $ & $20  $ & CSO	       &     \\
        $450 $ &  $6700   \pm 1300  $ & $54  $ & SCUBA	       & (2) \\
	$850 $ &  $600    \pm 120   $ & $54  $ & SCUBA	       & (2) \\ 
       $1110 $ &  $225    \pm 45    $ & $10  $ & SMA	       & (2) \\
       $1270 $ &  $240    \pm 20    $ & $54  $ & IRAM 30m      & (2) \\
       $1300 $ &  $190    \pm 30    $ & $10  $ & OVRO	       & (3) \\
       $2700 $ &  $20     \pm 2     $ & $10  $ & OVRO	       & (3) \\
    \hline
  \end{tabular}
  \caption{Photometric data points for CB~26. References: (1) \cite{s04}; (2) Launhardt et al., in prep.; (3) \cite{l01}}
\end{table*}

\section{Model \& Modelling}
In this section we provide the reader with an introduction to the concepts and techniques we use.

\subsection{The Model}
First, we discuss the various components of our model.
That is, the disc and envelope dust density distribution.
The employment of both these parts in the model is readily suggested by the data.

\subsubsection{The Disc}
The main part of our model for CB~26 is a parametrised   disc.
The disc is seen as the radially extended luminous structure in the millimetre maps.
The dark lane in the near-infrared maps is another indication for a disc as it can be understood as the disc's shadow on the surrounding envelope.

We employ a parametric approach to such a disc which can be written as
\begin{equation}
   \rho_\disc(\vec{r})= \rho_0 \left( \frac{R_*}{\rcyl} \right)^\alpha \exp \left(-\frac{1}{2}\left[ \frac{z}{h}\right]^2 \right)
   \label{eqdiscden}
\end{equation}
where $z$ is the usual cylindrical coordinate with $z=0$ corresponding to the disc midplane and $\rcyl$ is the radial distance from this $z$--axis \citep[see e.g.][]{w03a,s98}.
In our model, the parameter $\rho_0$ is determined by the mass of the entire disc.
$R_*$ is the stellar radius and $h$, the vertical scale height, is a function of $\rcyl$
\begin{equation}
   h(\rcyl) = h_0 \left( \frac{\rcyl}{R_*} \right)^\beta.
   \label{eqscalehight}
\end{equation}
Here the quantities $\alpha$, $\beta$, and $h_0$ in Equation (\ref{eqscalehight}) are geometrical parameters.
These parameters allow us to adjust the disc structure and shape in order to fit the data.
This modelling strategy has already been successfully applied to various other edge-on seen discs, such as the Butterfly-Star IRAS 04302+2247 \citep{w03a}, HK Tau \citep{s98}, IM Lupi \citep{p08}, and HV Tau \citep{s03}.

Integrating Equation (\ref{eqdiscden}) along the $z$ axis yields the surface density
\begin{equation}
  \Sigma(\vec{r})=\Sigma_0 \left( \frac{R_*}{\rcyl} \right)^p.
  \label{eqsurfaceden}
\end{equation}
Comparison with Equation (\ref{eqdiscden}) yields the following relation between
the exponent of the surface density power law and the geometrical parameters of the ansatz we use
\begin{equation}
  p = - \beta + \alpha.
  \label{eqp}
\end{equation}
The radial size of the disc $r_{\rm out}$ is another parameter and is mainly determined by the size of the elongated structure in the millimetre maps.
As \cite{h08} argues, the millimetre continuum does not trace the outermost region of the disc.
However, in the case of CB~26, the extent of the dark lane is also consistent with the outer radius estimate we get.

\subsubsection{The Envelope}
In order to reproduce the pattern of scattered light we see in the observations in the I, J, H, and K bands, we add an envelope-like dust distribution to the model.

The density within this envelope is understood to be orders of magnitude lower than that of the disc.
The HST optical and NIR data show while the envelope has a high enough density to produce scattered light, it is orders of magnitude lower than the density of the circumstellar disc which is optically thick enough to obscure the central star completely.
Further evidence for the low density in the envelope are the millimetre maps where the envelope is cannot be seen.

For the model of the envelope structure we follow the ideas of \cite{u76}.
We thus implement a model for a rotating envelope resulting from in-falling matter the same way as done by \cite{w03b,e05}:
\begin{equation}
  \rho_\env = \frac{\dot M}{8\pi\sqrt{GM_*r^3}} \left( 1+\frac{\mu}{\mu_0}\right)^{-\frac{1}{2}} \left( \frac{1}{2}\frac{\mu}{\mu_0} +\frac{r_{\rm cf}}{r}\mu_0^2\right)^{-1}.
  \label{eqenvrot}
\end{equation}
Here $\dot M$ is the dust in-fall rate, $M_*$ the stellar mass, $r_{\rm cf}$ the centrifugal radius and $\mu = \cos \theta$. The initial in-fall path of dust particles is given by $\mu_0$ as $r \rightarrow \infty$.
As this is the only occurrence of $\dot M$ and $M_*$, the factor $ \dot M ( 8\pi\sqrt{GM_*r^3})^{-1}$ in Equation \ref{eqenvrot} is merely a coupling constant that scales the mass of the envelope just as  $\rho_0$ does in case of Equation \ref{eqdiscden}.

Hence, we are using $ \dot M ( 8\pi\sqrt{GM_*r^3})^{-1} = \tilde{\rho}$ as a fitting parameter.
To avoid introducing a constant like $\tilde{\rho}$ with no direct physical interpretation into the discussion later on, we shall in this work refer to $M_*$ and $\dot M$ separately.

As a criterion to decide whether the disc dust distribution or the envelope dust distribution at a given point should be considered, we compare the two densities and choose the larger value:
\begin{equation}
\rho(\vec{r}) = \left\{
   \begin{array}{r@{\quad:\quad}l}
   \rho_\disc(\vec{r}) & \rho_\disc(\vec{r}) \ge \rho_\env(\vec{r})  \\
   \rho_\env(\vec{r})  & \rho_\disc(\vec{r})  <  \rho_\env(\vec{r})  \\
   \end{array} \right. .
   \label{dendiscenv}
\end{equation}
In this manner, we embed the disc into the envelope and guarantee a smooth transition from the disc to the envelope without the need to alter the density structure of the optically thick, millimetre-glowing part of the object.
For the radius of the complete model space we take twice the outer radius of the disc. 
Since we do employ a maximum size for the outer disc radius $r_{\rm out}$, the remaining space from the disc's edge to the end of the model space is readily filled by the envelope.
With a computational domain going out to $2r_{\rm out}$, we are able to model the scattered light from the envelope.

\subsubsection{The Dust}
Since gas is optically thin\footnote{
This implies that we neglect line emission and absorption by the gas.
We do not aim at modelling those as almost throughout the entire disc, the dust is by far the dominant coolant of the disc.
Hence, thermal equilibrium obtained in radiative transfer calculations based on dust only will provide a reliable description of the disc's thermal structure.
}  in the wavelength regime we deal with, we limit ourselves to radiative transfer through the dust.
For the mass relation of dust and gas we assume the standard value of
$  \frac{M_{\rm Gas}}{M_{\rm Dust}} = 100$
which is in agreement with the findings of \cite{g08} in another disc surrounding a low-mass T~Tauri star. 
Therefore, it is the dust whose density structure is described by Equations (\ref{eqdiscden}) and (\ref{eqenvrot}) in the disc and the envelope, respectively.
The dust grain properties in our model can be divided in three groups: The shape of the dust grains, their chemical composition, and their size distribution.

\paragraph{Grain shape}
We assume the dust grains to be homogeneous spheres.
Real dust grains, of course, are expected to feature a much more complex and fractal structure.
As discussed by \cite{v02}, chemical composition, size and shape of dust grains cannot be determined separately, but only as a combination.
We therefore limit our model to the less complex but also less ambiguous approach of spherical, non-aligned and non-orientated dust grains.

\paragraph{Grain chemistry}
For the chemical composition of the dust grains we employ a model that incorporates both silicate and graphite material.
This grain model has already been used to model the ``Butterfly star'' by \cite{w03a}.
For the optical data we use the complex refractive indices of ``smoothed astronomical silicate'' and graphite as published by \cite{w01}. 
Since the longest wavelength considered in our modelling is $2.7 \mm$, we extrapolate the refractive indices to that wavelength.
This is readily done since for this wavelength regime both the real and the imaginary part of the refractive index show asymptotic behaviour.
For graphite we adopt the common ``$\frac{1}{3} \ - \ \frac{2}{3}$'' approximation.
That means, if $Q_{\rm ext}$ is the extinction efficiency factor, then
\begin{equation}
   Q_{\rm ext, graph} = \frac{1}{3} Q_{\rm ext}(\epsilon_\parallel ) + \frac{2}{3} Q_{\rm ext}(\epsilon_\perp),
\end{equation}
where $\epsilon_\parallel$ and $\epsilon_\perp$ are the graphite dielectric tensor's components for the electric field parallel and orthogonal to the crystallographic axis, respectively.
As has been shown by \cite{d93}, this graphite model is sufficient for extinction curve modelling.
Applying an abundance ratio from silicate to graphite  of $\mathbf{1} \times 10^{-27} \cm^3 {\rm H}^{-1}$ : $\mathbf{1.69} \times 10^{-27} \cm^3 {\rm H}^{-1}$, we get relative abundances of $62.5 \%$ for astronomical silicate and $37.5\%$ graphite ($\frac{1}{3} \epsilon_\parallel$ and $\frac{2}{3}\epsilon_\perp$).

\paragraph{Grain sizes} 
For the grain size distribution we assume a power law of the form
\begin{equation}
  n(a) \mathbf{\dd a} \sim a^{-3.5}  \mathbf{\dd a} \quad {\rm with} \quad  a_{\rm min} < a < a_{\rm max}.
  \label{eqgraindist}
\end{equation}
Here, $a$ is the dust grain radius and $n(a)$ the number of dust grains with a specific radius.
For $a_{\rm min} = 5 \nm$ and  $a_{\rm max}=250 \nm$ this distribution becomes the commonly known MRN distribution of the interstellar medium by \cite{mrn}.
We choose those values as the starting point of the present study.

To model the different grain sizes and chemical populations, one has to consider an arbitrary number of separate dust grain sizes within a given interval $[a_{\rm min} : a_{\rm max} ]$.
But the observables derived from radiative transfer considering each grain species separately, are close to the observables resulting from radiative transfer (RT) simulations based on weighted mean dust grain parameters of the dust grain ensemble \citep{w03d}.
Thus, we use weighted mean values for the efficiencies factors, cross sections, albedo, and scattering matrix elements.
For each dust grain ensemble, 1000 logarithmically equidistantly distributed grain sizes within the interval $[a_{\rm min} : a_{\rm max} ]$ have been taken into account for each chemical component in the averaging process.

There are arguments that the grain size can not be governed by a power law as in Equation (\ref{eqgraindist}).
Furthermore, in the complex environment of a circumstellar disc, we expect dust settling and grain growth to make the grain size distribution quite dependent on the location within the disc. 
Unfortunately, a consideration of grain sizes that takes into account the effects of e.g. dust settling requires more than just one parameter as Equation (\ref{eqgraindist}) does. 
Given the data we have, we are not able to disentangle these parameters in our study.
Therefore, we assume that the power law distribution (\ref{eqgraindist}) to be valid in the whole disc and envelope structure and only use the maximum grain size as a parameter in our modelling efforts.

It will turn out that even with these simplifying assumptions we are able to model observational data of the system, although different $a_{\rm max}$ in the disc and envelope have been found -- but for other objects -- in the past \citep[e.g.][]{w03a}.

We assume an average grain mass density of $\rho_{\rm grain} = 2.5 \,{\rm g}\,{\rm cm}^{-3}$.
This density does not have any influence on the optical properties of the dust, as they are governed by the chemical composition and the particle size of the grains.
The average grain mass density merely controls, together with the disc mass and particle size, the number of dust grains in the disc.

\subsubsection{Heating Sources}
There are two sources of energy for the disc that need our attention.
The disc can be heated by stellar radiation and/or accretion of in-falling matter.

\paragraph{Accretion Heating}
Our model involves a parameter with the dimensionality of an ``accretion rate'': $\dot{M}$ in Equation (\ref{eqenvrot}) as part of the description of the envelope structure.
However, this quantity is really an in-fall rate within the envelope and may not necessarily at the same time describe mass accretion onto the star itself.
But it is in general the latter mass flow that, as e.g. in FU Orionis like objects, accounts for significant contributions to the system's luminosity.
For a T~Tauri like system as CB~26 we thus neglect accretion as a major source of energy.
As our study shows, $\dot{M}$ is rather small, around $\sim 10^{-8} \msun \, {\rm yr}^{-1}$.
This is small compared to FU Orionis objects but average for T~Tauri Stars and strongly supports our ansatz.

Besides matter in-fall in the envelope, accretion in the disc might be an important source of energy.
As shown by \cite{w03a}, the accretion luminosity from within the disc is about two of magnitude smaller than the stellar contribution.
As the model setup here is similar, we neglect accretion heating as a significant source of energy.

\paragraph{Stellar heating}
The discussion above leaves the star as the only primary source of energy in our model.
Its radiation heats the dust which then in turn itself re-emits at longer wavelengths.
In this sense the disc in our model is passive.
That is, we neglect accretion or turbulent processes within the disc as a possible other primary energy source.

We do not observe the star directly. This has mainly two reasons:
\begin{enumerate}
   \item In the far infrared and at longer wavelengths, where the disc becomes less opaque with increasing wavelength, the contribution to the spectral energy distribution of the dust is orders of magnitude larger than of the star.
   \item In the optical, near, and mid-infrared bands, the disc becomes opaque and the star is shielded from our direct view.
\end{enumerate}
Therefore, we have to assume the stellar parameters, that is temperature and luminosity, since we are not able to derive them directly from observations.
Observations only hint at a luminosity being $L \ge 0.5 \lsun $.
As a starting point we choose an ``average'' T~Tauri star as described by \cite{g88}.
This star has a radius of $r=2\rsun$ and a luminosity of  $L_* = 0.92 {\rm L}_\odot$.
Assuming the star to be a black body radiator, this yields an effective surface temperature of $\teff=4000 \K$ .

Both of these parameters have been kept fixed in our parameter study to avoid degeneracies between parameters of the model. Except for the total flux, this choice has no impact on the near-infrared images.

As \cite{n93} showed, under certain conditions stellar light scattered back to the disc can have significant implications for the thermal structure of the disc.
Here, the outer envelope regions $400\au \le r_\env \le 1000\au$ is shown to be of importance.
An important assumption in the argumentation is the scattering phase function to be independent of the scattering angle.
However, in our modelling framework the scattering phase function is highly asymmetrical and favours forward scattering by orders of magnitude.
Thus, the amount of radiation scattered back to the disc from the envelope outside our model space, i. e. at distance larger then $400\au$ can be neglected.

\subsection{Means of modelling}
In this section we discuss how we proceed with the aforesaid model.
We discuss the free parameters of the model, their range, and the sampling of the resulting parameter space.
Finally, we review the constraints imposed upon our model by the various observation and give a criterion for the best-fit model.

\subsubsection{Radiative Transfer}
For our continuum radiative transfer simulations we made use of the program \texttt{MC3D} \citep{w99,w03c}.
It is based on the Monte-Carlo method and solves the continuum radiative transfer problem self-consistently.
It estimates the dust temperature distribution taking into account any heating sources, in our case the central star's radiation.
It makes use of the temperature correction technique as described by \cite{b01}, the absorption concept as introduced by \cite{l99} and the enforced scattering scheme as proposed by \cite{c59}.
The optical properties of the dust grains (scattering, extinction and absorption cross sections, scattering phase function) and their interaction with the radiation field is calculated using Mie theory.
Multiple and anisotropic scattering is considered.
The phase function is highly asymmetrical (e. g. at the peak of stellar emission at $\lambda=0.7\mum$ one has $\left <\cos(\theta_{\rm scatter})\right> = 0.86$), strongly favouring forward-scattering.

In order to derive a spatially resolved dust temperature distribution, the model space has to be subdivided into volume elements inside which a constant temperature is assumed.
Both the symmetry of the density distribution and the density gradient distribution have to be taken into account.
For the present study, we use a spherical model space, centred on the illuminating star and an equidistant subdivision of the model in the $\theta$-direction, whilst a logarithmic radial scale is chosen in order to resolve the temperature gradient at the very dense inner region of the disc.
The required spatial resolution at the disc inner radius rim of our model ranges from $10^{-4}\au$ up to $10^{-1}\au$ and every grid cell outwards is 1\% larger than its next inner neighbour.

The radiative transfer is simulated at 101 wavelengths.
The first 100 wavelengths are logarithmically distributed in the wavelength range  $[\lambda_{\rm min},\lambda_{{\rm max}-1}] = [50 \nm, 2.0 \mm ]$.
The largest wavelength used is $\lambda_{{\rm max}} = 2.7 \mm$.

With \texttt{MC3D} we compute observables from the model.
These observables are then compared to the observed data in the quest for the best-fit model.
Namely, the quantities we derive with \texttt{MC3D} from the model are
\begin{enumerate}
  \item Images in the NIR, that is in the I, J, H, and K Band,
  \item Images in the millimetre regime at 1.1\,mm, 1.3\,mm and 2.7\,mm,
  \item 101 points for the SED accordingly to the above wavelength distribution.
\end{enumerate}

\subsubsection{Constraints from observations}
Facing the broad variety of available observational data, one has to point out what the main features are that we want to reproduce with our model.
This also determines the criteria for the best-fit model.
The case is simple for the spectral energy distribution.
There we aim at reproducing the complete spectrum over three orders of magnitudes from the optical bands down to the millimetre regime.
We can divide the maps of the disc in two major groups: the maps in the millimetre regime, and the maps in the near-infrared.
Both groups trace different physical processes and different spatial regions of the object.

\paragraph{Millimetre maps}
For resolved images, the issue is not as simple as for the SED.
Our model is rotational symmetric and thus does not provide for any related asymmetry as seen in observations.

The morphology of the millimetre maps has its origin in the dust that is heated by the star and re-emits light at those wavelengths.
Although the images at $1.1 \mm$, $1.3 \mm$ and $2.7 \mm$ are rather simply structured they impose two major features that constrain our models.
These are
\begin{enumerate}
   \item the peak flux and
   \item the spatial brightness distribution.
\end{enumerate}
Since in all three maps the beam size is larger or comparable to the vertical extent of the disc, we can not constrain the flux distribution on the $z$--axis,  perpendicular to the disc mid plane.
Any feature there is smoothed out by the beam.
Therefore, we focus on reproducing the flux distribution along the midplane of the disc.
All images are fitted in the image plane.

\paragraph{Maps in the near-infrared}
The four images in the near-infrared show more structures and details than the millimetre maps.
Besides the disc appearing as dark lane in the near-infrared also a complex, wavelength-dependent morphology of the surrounding envelope is seen.

Considering that the circumstellar disc CB~26 is located at the edge of a Bok globule, one realises that the environment of the disc can account for the majority of the substructure seen on these maps, yet the Bok Globule itself is not part of our model.
Hence, we have to restrict ourselves to the following two points that we want to reproduce with our model:
\begin{enumerate}
   \item the dependence of the width of the dark dust lane on wavelength and
   \item the relative peak height of the brightness distribution above and below the dust lane.
\end{enumerate}
We restrict our modelling efforts to a simple envelope structure and the above two points since it is hard to distinguish whether the appearance of the object in the observations is due to envelope or environmental structure.
We also put no emphasis on the vertical width of the upper and lower lobe nor the exact morphology.

\subsubsection{Quality of the fit}

For each comparison between model and observation on the aforesaid points, we get an individual $\chi^2_i$.
The total ${\chi}^2$ of one model is then just the sum over all the individual ${\chi}^2$s:
\begin{equation}
   {\chi}^2_{\rm total} = \frac{1}{n}\left({\chi}^2_{\rm SED} + \sum_{\rm mm-maps} {\chi}^2_i + \sum_{\rm NIR-maps} {\chi}^2_j\right)
  \label{eqsumxi}
\end{equation}
Here, $n=8$ as we have one ${\chi}^2$ from the SED, three from the millimetre maps and four from the scattered light images.

Based on ${\chi}^2_{\rm total}$ we get from Equation (\ref{eqsumxi}), we give our modelling errors as the range where we can alter the parameter values without changing ${\chi}^2_{\rm total}$ more than $10\%$.
This value is rather arbitrary as there is no mathematical reasoning behind it.
Yet, it has proved within our study to reflect quite well the adjustability of the model.
Allowing for a larger variation of $\chi^2_{\rm total}$ than $10\%$ gives generally worse results.

\subsubsection{Parameter space study}
Based on the model outlaid in the previous sections, we are left with ten adjustable parameters to reproduce the characteristics as described in the previous section.
If not stated otherwise, we choose the range of a parameter for our study based on modelling of other yet similar objects \citep[see e.g.][]{w03a}.
Then we first sample that range of an individual parameter in four coarse steps, select the two best values and go with the same procedure to the next parameter.
Secondly, we take the results as an indicator how to refine the stepping in which smaller range.
This process is being iterated to reach our final results.
As the starting values in our parameter space study we chose the values obtain for circumstellar disc IRAS 04302+2247 which has at first glance a similar appearance as the disc in CB~26.

In detail the parameters we have are
\begin{enumerate}
   \item The exponents $\alpha$ and $\beta$ which describe the radial density profile of the disc (see Equation (\ref{eqdiscden}) for details).
	From \cite{d99}, we choose for the flaring parameter $\beta=1.25$ and get then a corresponding value for $\alpha=2.25$ from the relation $\alpha = 3 (\beta-\frac{1}{2})$, which is a result of accretion disc physics (see e.g. \cite{s73}). 
Those values are taken as a starting point and we then look for agreement between observations and modelling at and beyond those values in steps of $0.1$ and $0.2$, respectively.
  \item The scale height $h_0$ of the disc at a given radius.
      In the following we will fix the scale height at a radial distance from the star of $\rcyl=100\au$ and consider the cases $h_0=5\au$, $10\au$, $15\au$, $20\au$, and $25\au$.
  \item From the density Equation (\ref{eqenvrot}) for the envelope we have the centrifugal radius $r_{\rm cf}$ and
  \item the coupling constant $\tilde{\rho}$ which is a function of the mass accretion rate $\dot{M}$ and the mass of the central star $M_*$.
	For the centrifugal radius we probe values in the range $100 \au \le r_{\rm cf}\le 800 \au$.
	As for $\tilde{\rho}$, we chose for the accretion rate values from the interval $\dot{M}~\in~[10^{-5}\msun {\rm yr}^{-1};10^{-10}\msun {\rm yr}^{-1}]$ and calculate $\tilde{\rho}$ under the consideration of the stellar mass by \cite{l08}. 
	In this paper a dynamical mass of $M_* =0.5 \pm 0.1 \msun $ is derived.
  \item The inner and outer disc radius.
	The inner radius was initially set to $0.1\au$, which is approximately the dust sublimation radius.
	However, we do not get good agreement especially with the $1.3 \mm$ map unless we choose values of $\sim 45\au$.
	For the outer radius, we chose a value of $200 \au$ but also considered configurations with $ 150 \au$ and $250 \au$.
  \item The disc mass.
	We consider 7 different disc dust masses in the range from $1.0\times10^{-5}~\msun$ up to $4.5 \times 10^{-3} \msun$.
  \item The disc's inclination is such that the system is seen almost edge-on.
	We restrict the range for $\theta$ to be part of our parameter space to values between $60^\circ$ and $90^\circ$ with a stepping of $1^\circ$ between $80^\circ$ and $90^\circ$.
  \item The maximum grain size $a_{\rm max}$.
	Grain growth is a major issue for protoplanetary discs  as it is the first step towards the formation of planets from the dust in the interstellar medium.
     We allow for a maximum grain size of $1 \mm$.
\end{enumerate}

\begin{figure}
  \resizebox{\hsize}{!}{
    \includegraphics{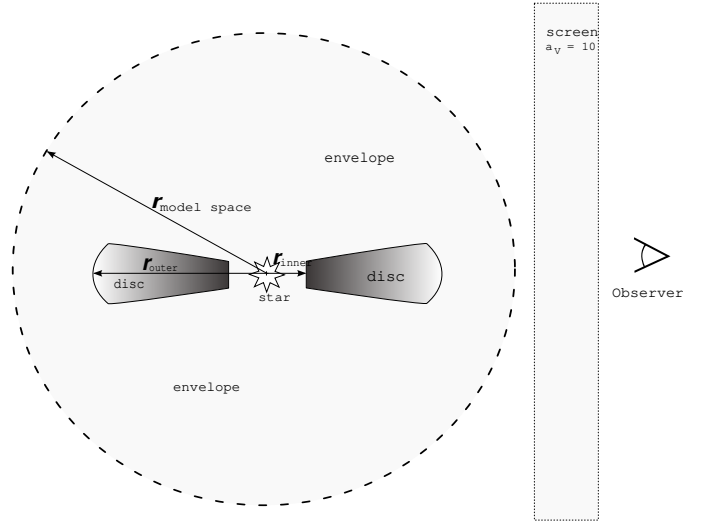}
  }
  \caption{
    Sketch showing all components of our model and their spatial arrangement.
    All sizes are not to scale.
  }
  \label{modelfig}
\end{figure}

For a summary of those ten adjustable parameters of our model and the range we covered in the study see Table \ref{Table:1}.
A sketch showing all components of our model is presented in Fig. \ref{modelfig}.

\section{Results}
\begin{table*}
  \centering
  \begin{tabular}{c c c c c}
      \hline \hline
      Parameter & Minimum value & Maximum value & Best fit value & Uncertainty \\
     \hline 
      $\alpha$			  & $2.0$		  & $5.0$   & $2.2$	& $\pm 0.1$ \\
      $\beta$			  & $1.0$		  & $2.6$   & $1.4$	& $\pm 0.1$ \\
      $h_0 [{\rm AU}]$		  & $5 $		  & $25$    & $10$	& $\pm 2.5$ \\
      $r_{\rm in}\, [{\rm AU}]$	  & $0.1 $		  & $60$    & $45$	& $\pm 5$ \\
      $r_{\rm out}\, [{\rm AU}]$  & $150 $		  & $350$   & $200$	& $\pm 25$ \\
     \hline 
      Dust mass [$\msun$]	      & $1.0 \times 10^{-5} $ & $5.0 \times 10^{-3} $ & $3.0 \times 10^{-3} $ & $\pm 0.2 \times 10^{-3} $ \\
      $\dot{M} [\msun {\rm yr}^{-1}]$ & $10^{-10} $	      & $10^{-5} $	      & $10^{-8} $	      & $\pm 0.5 \times 10^{-8} $ \\
      $r_{\rm cf}  [\au]$	      & $100$		      & $800$		      & $460$		      & $\pm 10$ \\
      $a_{\rm max}\, [\mum]$	      & $0.25$		      & $1000$		      & $2.5$		      & $\pm 0.3$ \\ 
     \hline
      $\theta$ & $60^\circ$ & $90^\circ$ & $85^\circ$ & $\pm5^\circ$ \\
      \hline
  \end{tabular}
  \caption{Overview of parameter ranges and best-fit values.
    For the definition of the uncertainty see section 3.2.3.
    The first group of parameters contains purely geometric parameters, the second group physical parameters and the last group the inclination of the disc as seen by the observer as an observational parameter.}
   \label{Table:1}
\end{table*}

The values of the parameters of our best-fit model can be found in Table \ref{Table:1}.
Our geometrical parameters $\alpha$ and $\beta$ of the disc density structure yield with the Equation (\ref{eqp}) a surface density power-law exponent of $p=-0.8$.
\begin{figure}
  \resizebox{\hsize}{!}{
    \includegraphics{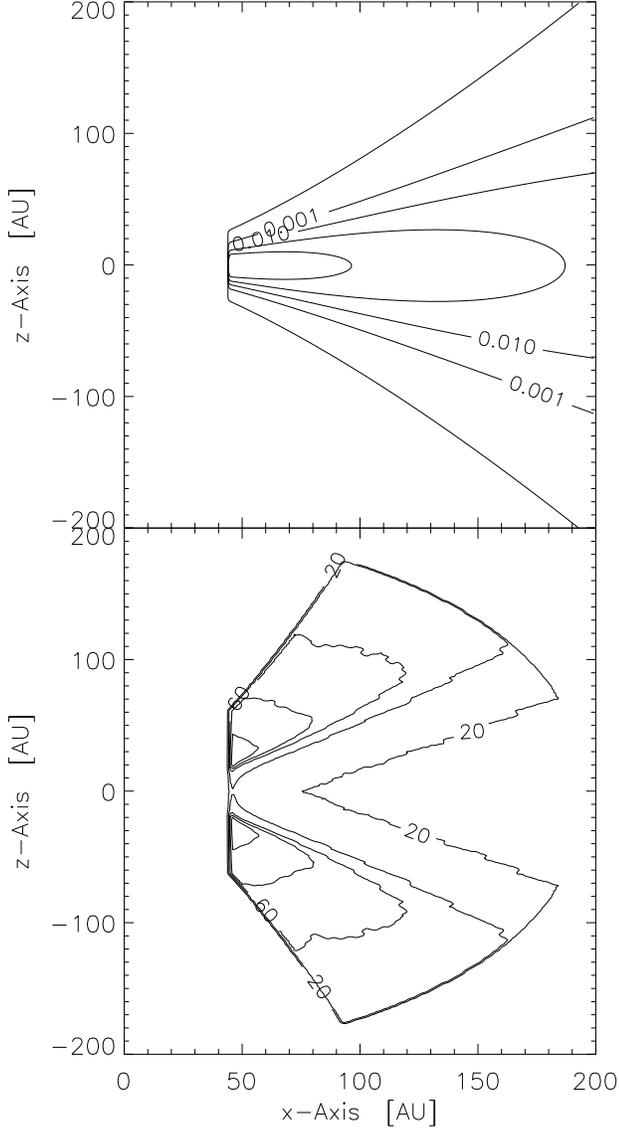}
  }
  \caption{
    \emph{Upper plot:} Contours of dust density distribution on a plane perpendicular to the midplane normalised to the peak density of $0.34 \times 10^{-8} \,{\rm g}\,{\rm cm}^{-3}$.
    The contour levels are at $10^{-7}$, $10^{-3}$,$10^{-2}$, $5\times10^{-2}$, and $2\times 10^{-1}$. 
    \emph{Lower plot:} Contours of the temperature distribution on the same plane as above.
    Contour levels are at 20K, 40K, 60K, 70K, and 80K.
    The maximum temperature is 90K.
  }
  \label{dentemp}
\end{figure}
\begin{figure}
  \resizebox{\hsize}{!}{
  \includegraphics{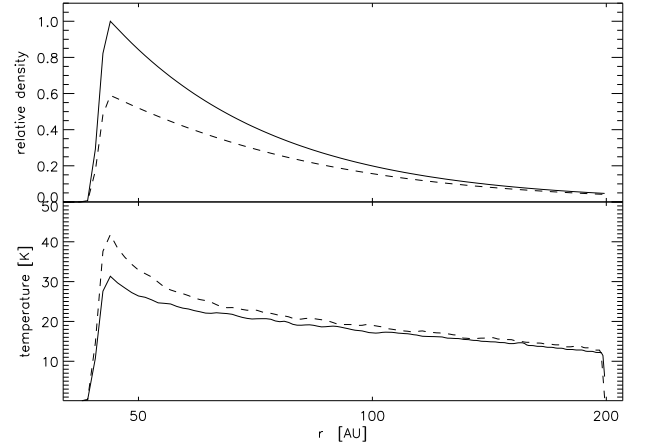}
  }
  \caption{
    Radial profiles of density (\emph{upper plot}) and temperature (\emph{lower plot}) in the midplane along $r$ (\emph{solid line}) and along $z/r=0.1$ (\emph{dashed line}). The latter profile is chosen in order to include the scale height $h_0$ at $r=100\au$.
    The maximum density is normalised to $1$ (corresponding to $0.34 \times 10^{-8} \,{\rm g}\,{\rm cm}^{-3}$).
  }
  \label{radprofs}
\end{figure}

In Fig. \ref{dentemp} the density and temperature distribution of the best-fit model is shown while in Fig. \ref{radprofs} radial profiles of the density and temperature distribution in the midplane and $20 \au$ above are shown.
One can see, that the high density in the midplane on the inner rim provides enough opacity such that material behind it in the midplane is not being heated up directly by the stellar radiation.
The highest dust temperature at the inner disc rim amounts to $\approx 90 {\rm K}$ and is reached $\sim 20 \au$ above the midplane.
However, due to the high density in the midplane, the stellar radiation does not penetrate deep into the midplane which results in the steep temperature gradient.
In the less dense upper (and lower) layers of the disc, the stellar radiation also heats more distant parts of the disc resulting in a less steep temperature gradient.

An average dust temperature of $\bar{T}_{\rm dust}=16\K$ is obtained from the temperature distribution by weighting it with the mass distribution.
This goes nicely with Fig. \ref{dentemp} if one bears in mind that the bulk of the dust is located in the midplane and well shielded against stellar radiation by inner parts of the disc. 
High temperatures are only reached in a very narrow region at the inner disc rim and in the very low density regions of the disc and thus contributing little to the mass averaged dust temperature.

Since our dust grain model only uses refractive indices, an effective dust grain opacity can be calculated by re-arranging
\begin{equation}
  M_{\rm dust} = \frac{S_\nu D^2}{\kappa_\nu B_\nu(\bar{T}_{\rm dust})}
  \label{eqop}
\end{equation}
and assuming gas-to-dust ratio of 100.
Here, $\kappa_\nu$ is the wavelength depended mass absorption coefficient, $S_\nu$ is the observed flux, $D$ the distance of the object, and $B_\nu(\bar{T}_{\rm dust})$  the Planck-function at a certain temperature.
The calculation yields for our model a dust opacity of $\kappa_{1.3\mm}=0.26 \,{\rm cm}^2{\rm g}^{-1}$.
This value is very close to the ISM dust opacity given by \citet{d84,d87}.
Compared to opacities for coagulated dust grains and ice-coated grains \cite{o94,B90}, our model yields an effective dust grain opacity at the lower end of the range of commonly employed opacities.

Fig. \ref{SED} shows the spectral energy distribution of the best-fit model in comparison with the spectral data.
We achieve quite a good match to the observational data except for the optical wavelengths.
Since the circumstellar disc is embedded in the Bok globule CB~26 we need to care about the dust outside our model space as well.
Thus, we add a screen that mimics the effect of foreground extinction between the object and the observer.
For this screen, we assume the extinction properties of interstellar dust grains.
Such a screen is described in detail by \cite{c89}.
Using $A_{\rm V}$ as a parameter with a minimum value of 2, we find in our study that a screen with a visual extinction of $A_{\rm V} = 10$ can easily account for the missing flux in the optical.
The result is shown in Fig. \ref{SED} as the dashed curve, whereas the dashed-dotted line corresponds to the best-fit model without the screen.

It needs to be stressed, that this screen only applies extinction law to all observable quantities.
It is not subject to any radiative transfer or thermal re-emission.
We estimated the possible contribution to re-emitted radiation of a such a screen with $A_{\rm V} = 10$ composed of ISM grains at the same distance as CB~26 with a temperature of 16\,K.
We found the screen to be clearly optically thin (e.g. $\tau_{1.3\mm} = 8\times 10^{-5}$) and has only enough mass to have about $\approx 1\%$ of the observed flux in the millimetre regime.

\begin{figure}
  \resizebox{\hsize}{!}{
    \includegraphics{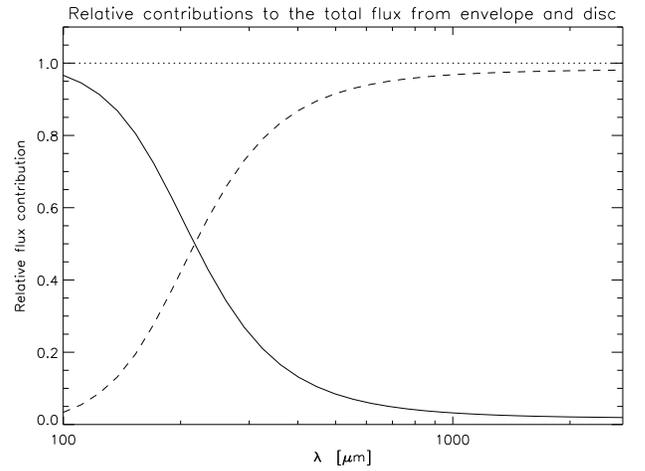}
  }
  \caption{
    Contributions to the spectral energy distribution from disc (dashed line) and envelope (solid line).
    The transition from envelope to disc as the major source of radiation (re-emission and scattering) is at $\lambda=217\mum$.
  }
  \label{relsed}
\end{figure}

Furthermore, the spectral energy distribution of the model shows that the contribution of the envelope is quite important for shorter wavelengths.
In this regime, the main contribution to the spectral energy distribution comes from the envelope whilst in the radio regime the flux comes completely from the dust in the disc that glows at those wavelengths.
Fig. \ref{relsed} illustrates this.

\section{Discussion}
The following paragraphs are now dedicated for a more detailed discussion of some results of our model.

\subsection{Grain size and growth}
It needs to be pointed out, that we found a model capable of explaining all major elements  of the observations without the need to increase the maximum grain size in our parameter study significantly.
The maximum grain size of our best fit model is $a_{\rm max} = 2.5 \mum$.
While this is a factor ten larger than the smallest maximum grain size considered in our parameter space, the value found is only marginally larger then upper grain sizes given for the ISM in the literature.
In fact, this is only true for the dust in the the disc component of our model.
The maximum grain size in the envelope is the same as in found in the ISM, $a_{\rm max} = 0.25 \mum$.
If the maximum grain size in the envelope were bigger, then the short wavelength part of the SED could not be reproduced.

It is certainly smaller by several orders of magnitude than the maximum grain size found in other disc models such as in the work of \cite{p08}.
There, a maximum grain size of a few millimetres has been found.
In contrast, models in our parameter space featuring values of $a_{\rm max} \approx 1\mm$ fail to fit the SED in the millimetre regime as the slope of the model SED is not steep enough.
Also, we did not succeed to reproduce the default value for the maximum grain size of $a_{\rm max}=250\nm$ of ISM.
In particular, the model would be off by a factor of ten for the SED data point at $1.3 \mm$.

One needs to discuss why the slight change for $a_{\rm max}$ from $250\nm$ to $2.5\mum$ allows for a fit in the \emph{millimetre} part of the SED -- at wavelengths three orders of magnitude larger than the largest grains in the model.
Intuitively, one would expect the millimetre part of the SED to remain unaltered by a change of grain size at that level.
However, this expectation is based on the assumption that the absorption efficiency of the grains $C_{\rm abs}$ in the millimetre regime is also insensitive to a change in the grain size at the micrometre level.
Yet, this only holds true for only two of the three dust species in the model.
For the astronomical silicate and the graphite component with an alignment of the crystals' optical axis perpendicular to the propagation direction of the electromagnetic field  $C_{\rm abs}$ has the same slope in the millimetre regime for grain size distributions with a maximum grain size of $250\nm$ and $2.5\mum$.
But for the third dust species, namely the graphite component with the crystals' optical axis aligned with the electromagnetic field, this is different.
Here, the slope of  $C_{\rm abs}$ is significantly larger for  $a_{\rm max}=250 \nm$ than for  $a_{\rm max}=2.5 \mum$.
This effect is large enough to dominate the sensitivity of the SED to changes in the maximum grain size even at the level discussed despite the fact that the dust species responsible for this behaviour has only a $12.5\%$ share of the total dust.
This is due to the fact that graphite is a far more effective absorber than silicate.

A look on the millimetre spectral index of the data yields $\alpha_{\rm mm} = 3.1\pm.27$.
The corresponding millimetre opacity slope is $\beta_{\rm mm} = 1.1\pm.27$, if the millimetre emission is assumed to be optically thin.
A $\beta_{\rm mm} = 1$ and smaller is understood to indicate dust grain particles larger than in the interstellar medium to be present.
A value of $\beta_{\rm mm} = 2$ is expected if only ISM grains were present in the disc.
The latter is true only for grains whose absorption efficiency  $C_{\rm abs}$ behaves like silicate.
The value of $\beta_{\rm mm}$ we obtain from data and model of CB~26 close to what is expected for large grains despite having still only micrometre sized grains in the model is due to the unorthodox behaviour of the parallel graphite component on the one hand side and on the other hand side to the non-vanishing optical depth in the millimetre regime (e.g.  $\tau_{1.3\mm}\sim 0.6$).

\cite{d06} considered also the behaviour of the millimetre opacity slope $\beta_{\rm mm}$ for dust mixtures of graphite and silicate.
There, no dependency of the opacity index as in our work was found.
However, in this analysis not crystalline graphite was used but the optical properties of amorphous carbonaceous solids instead.

We summarise that we do not need mm sized grains to model the circumstellar disc CB~26 but grains with a maximum grain size still close to what is found in the ISM.
This is in contrast to the modelling of the Butterfly star \citep[see][]{w03a} as well as for the circumstellar disc HH~30 \citep[see][]{w02} where in both cases the authors found it necessary, to have their largest grains at least four times larger than the largest grain of the interstellar matter.
Of course, this result is based directly on the choice of the grain model we made.
For another model, especially one without graphite, larger grains might be needed to fit the observed SED.
Yet, due to the poorly constrained dust composition of circumstellar discs -- in particular in the disc interior -- this degeneracy between dust model and grain size must remain.
As the dust model used in this work is also used in the context of other studies of circumstellar discs such as the Butterfly star \citep{w03a}, it is a reasonable choice as it keeps the models of similar objects comparable as they are built on common assumptions.

\subsection{Inner hole}
\begin{figure}
   \resizebox{\hsize}{!}{
    \includegraphics{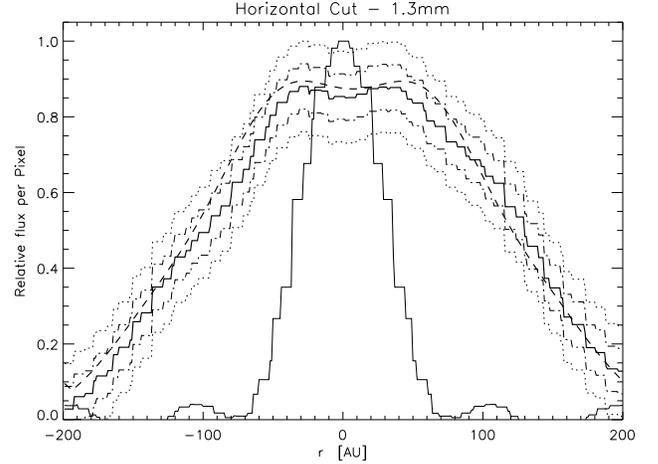}
   }
   \caption{Horizontal cut through the spatial brightness distribution at $1.3 \mm$.
      The thick solid line represents the OVRO observation,
      the dash-dotted lines gives the addition/subtraction of one $\sigma$,
      the dotted line indicates the $2\sigma$-levels,
      the dashed line corresponds to our best-fit model,
      the thin solid is the dirty beam of the observation.}
   \label{radiocut}
\end{figure}

The most unexpected result of our modelling is the inner disc radius for the model.
In Fig. \ref{radiocut} the solid line shows the flux profile along the disc midplane at $1.3 \mm$ as seen by OVRO whilst the thin solid line is the PSF of the observation.
At the centre of the disc, the profile shows a plateau in the brightness distribution.
There are two explanations in stock for this dip.

First, the minimum could indicate that emitting dust in that region is present but not visible.
If the optical depth in the midplane is sufficiently high, the flux contribution from the inner parts of the disc compared to the contribution of the optically thin parts on the disc's surface would be smaller. 

A high optical depth can easily be reached in regions of high dust densities.
For a given total disc mass, a disc with small inner radius yields higher densities than the model with the large inner radius.
Hence, the smaller the inner radius in our model, the more matter we find to be close to the star and thus reaching higher optical depths in the inner disc regions.
However, even for the smallest inner radius of our parameter space, $0.1\au$ we did not reach an optical depth that obscures enough flux from the disc centre.

This behaviour would also be more obvious when compared with maps at shorter wavelengths since the optical depth increases with decreasing wavelength.
But in the images at $1.1 \mm$ and $2.7 \mm$ we do not observe a dip at the centre of the disc.
Unfortunately, the available images do not help us to conclude whether the absence of the dip is really an indicator for a big void in the disc.
This is because the point spread function at those two images is far too large to resolve the feature (see Table \ref{beamTable}).
\cite{w08} reasoned this way in the case of IRAS 04302+2247 where they found a similar dip in the brightness distribution at $\lambda=894\mum$ but not at $1.3 \mm$. 

The second possible cause for the spatial brightness distribution in the $1.3 \mm$ map is the actual lack of dust in the inner region of the disc.
Whilst we started with an inner radius in the order of magnitude of a few tens of the stellar radius, it was not possible to match the plateau structure of the $1.3 \mm$ map. 
On this account, we allow the inner disc radius to be as large as $50 \au$.
Unfortunately, within our model and parameter framework we are not at liberty to increase the total mass, and hence the density at the disc centre because
 the flux at $2.7 \mm$ sets already an upper limit for the total disc dust mass.
 This is because at this wavelength any model is optically thin and thus we see the total matter of the disc.

Another way to think about the issue is to consider the optical thickness of the disc at $1.3 \mm$.
We have run one simulation of the disc with exactly the same parameter values as for our best-fit model except for the inner radius. For the comparison model we chose $r_{\rm in}=0.1\au$.
For both runs we computed the optical depth along the line of sight from the observer through the disc.
As a result, for the large inner radius we have $\tau_{1.3\mm} = 0.6$ and for the small inner radius $\tau_{1.3\mm} = 1.9$.
In the first case we deal with an optically thin system.
As for the second, this is not so clearly said.
An optical depth of $1.9$ means that the initial flux is reduced by a factor of $e^{-1.9}=0.15$.
A much larger value of $\tau$ along the line of sight would be required to hide all emitting dust in the central region and produce the observed plateau structure.
In our study, we were not able to fit the millimetre profile with a small inner radius.

We are therefore forced to conclude, that in the millimetre regime we do see the entire disc. 
Hence, the plateau structure rises from the wide spatial separation of the inner rim from the star, whereas a disc with a small inner radius would have a central peak in the brightness distribution.

\begin{figure}
  \resizebox{\hsize}{!}{
    \includegraphics[width=0.5\textwidth]{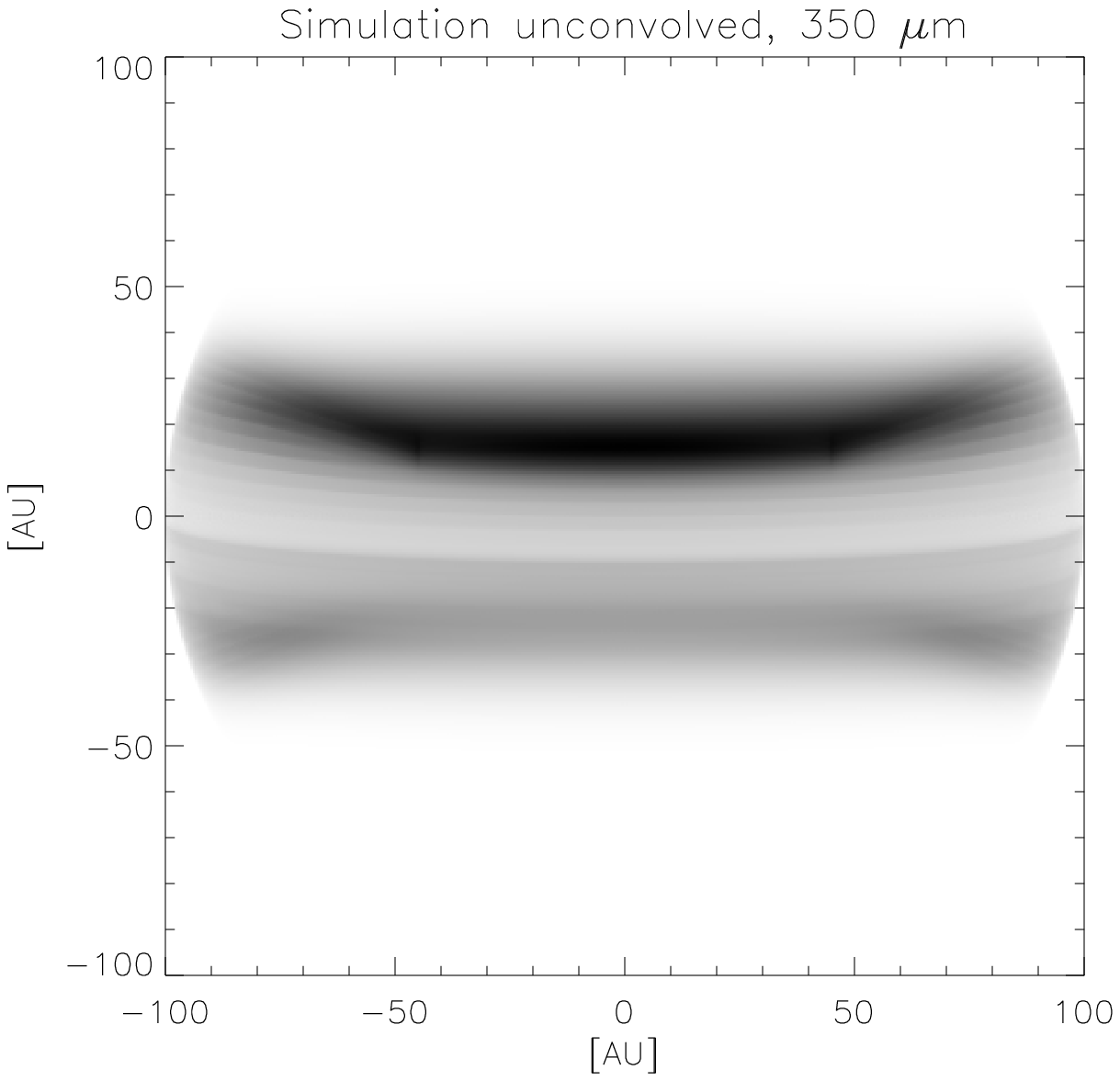}
    \includegraphics[width=0.5\textwidth]{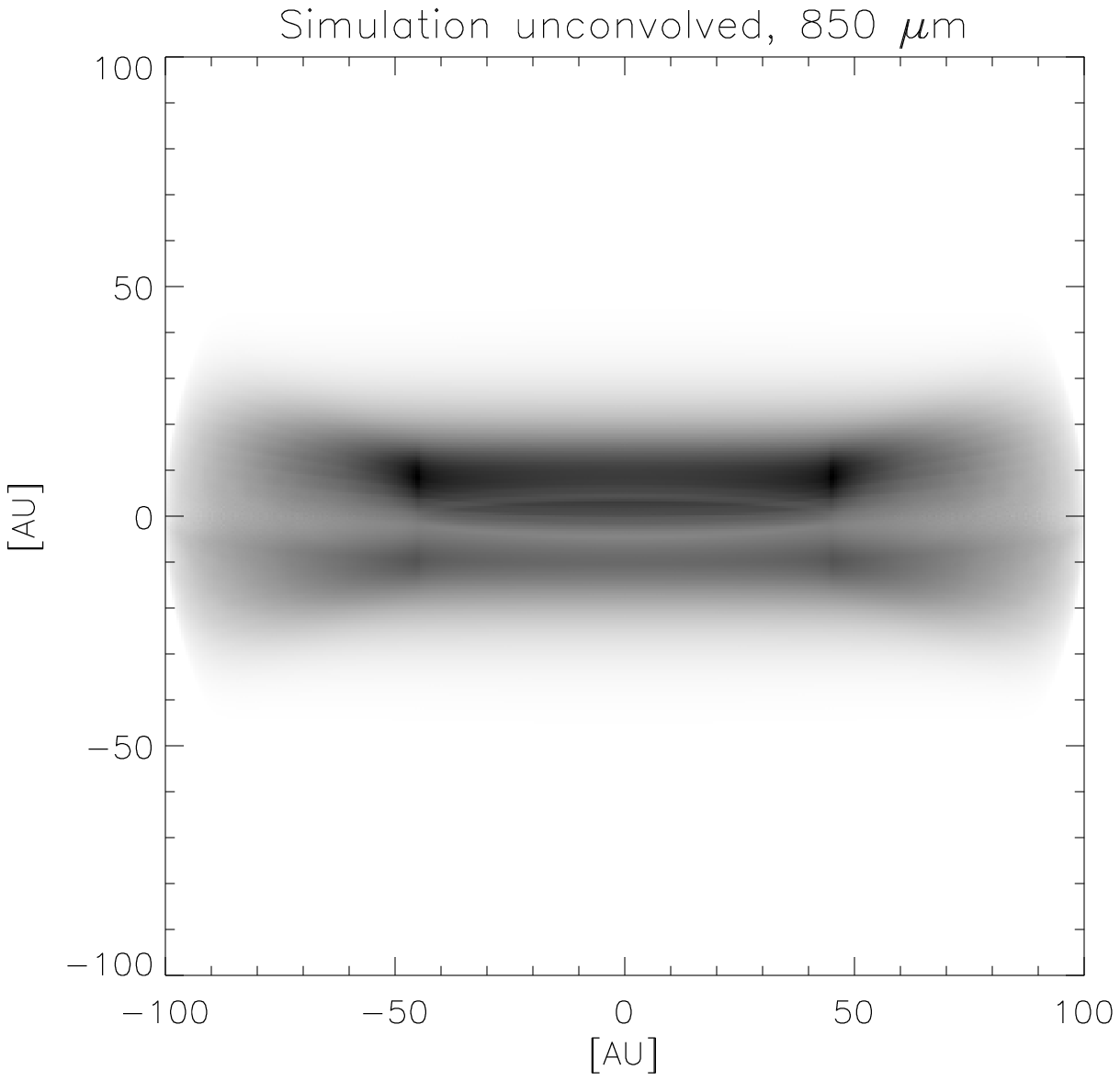}
  }

  \resizebox{\hsize}{!}{
   \includegraphics[width=0.5\textwidth]{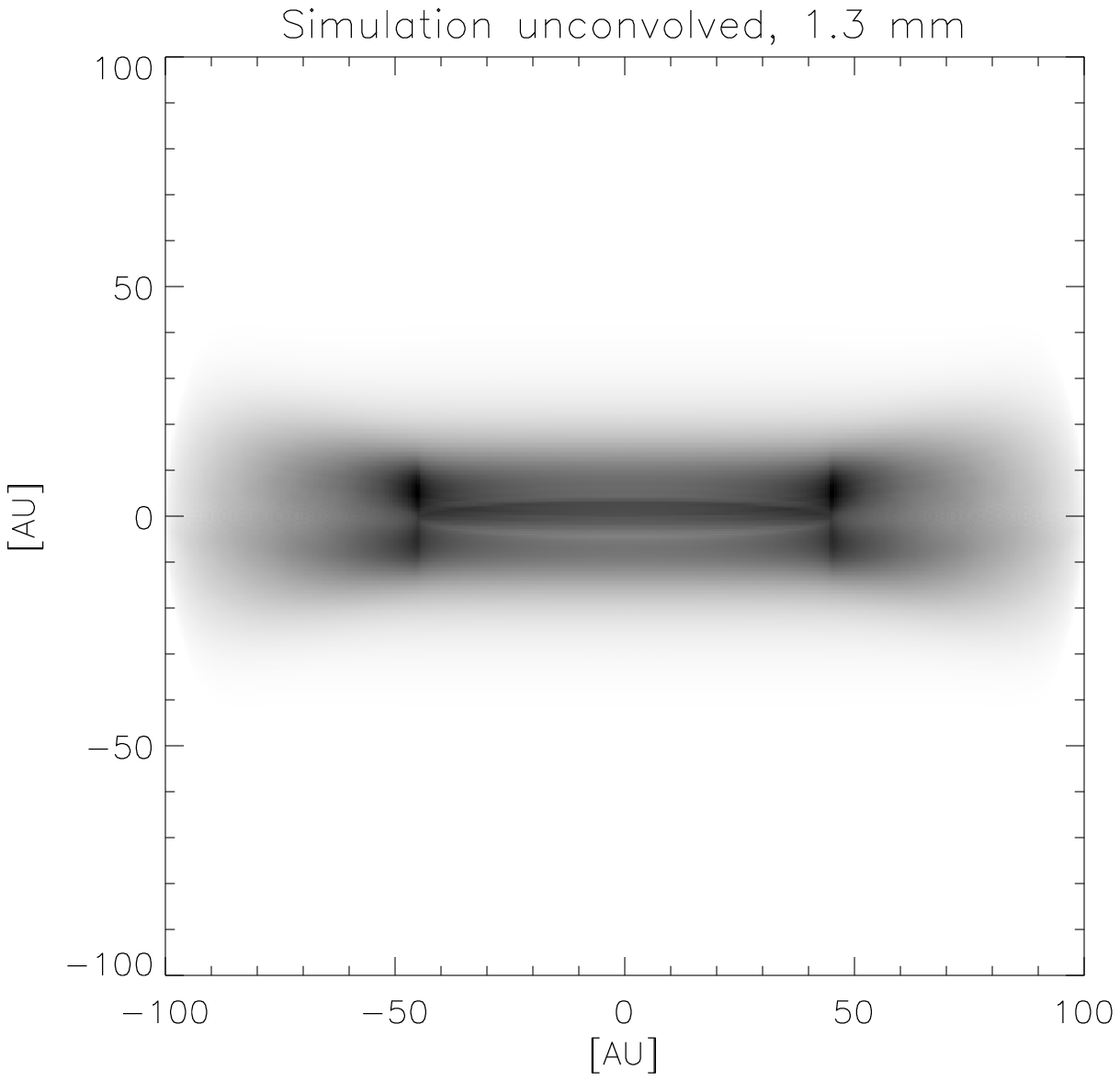} \\
   \includegraphics[width=0.5\textwidth]{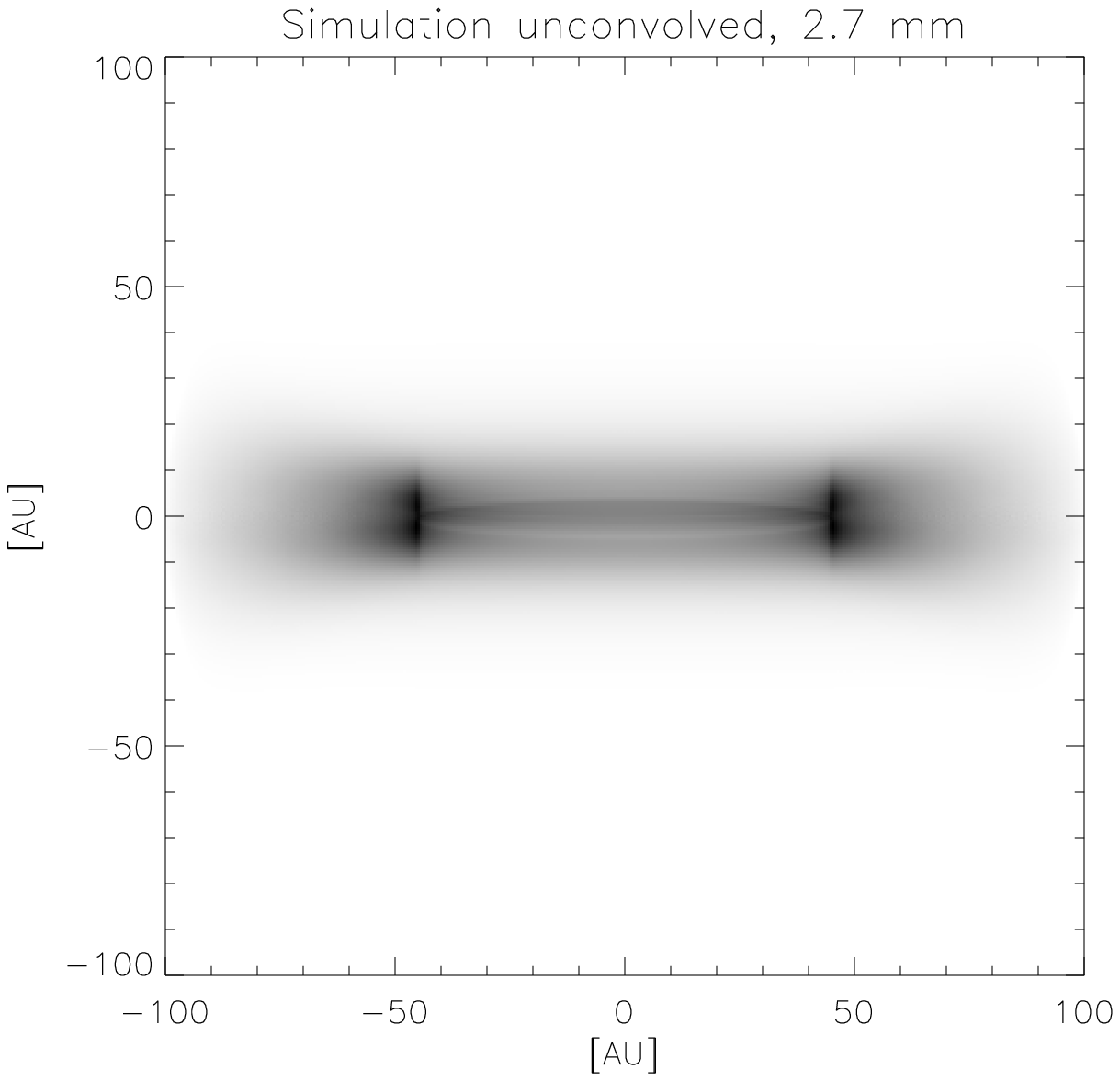} \\
  }
   \caption{Predicted appearance of the inner region of the disc.
   }
   \label{prediction}
\end{figure}

Based on this line of arguments, our model provides predictive power for high resolution images at wavelengths longer than $1.3 \mm$. 
Since we conclude that the plateau structure in the brightness profile is due to lack of dust, it should also be visible at longer wavelengths.
There, the dust becomes ever optical thinner and thus provides us with literal insight inside the disc.
Now, if the lack of flux in the disc centre would be due to an effect of optical depth, it should vanish with longer wavelengths and the brightness profile should have one central peak instead of a plateau-structure.
Unfortunately, the image at $2.7 \mm$ has a point spread function far too large to allow us to disentangle between the two possibilities.
Future observations of the disc CB~26 with a high spatial resolution will provide a perfect opportunity to confirm this prediction.
Fig. \ref{prediction} shows what we expect the disc to look like at different wavelengths according to our model with the inner void.
With the Atacama Large Millimeter Array (ALMA) it will be quite easy to confirm our findings.

It needs to be pointed out, that the spectral energy distribution from CB~26 itself does not hint on the large inner hole.
Its general shape is similar to other edge-on seen disc's.
This is because the inner regions of the disc, while they completely dominate the disc emission in the 1-20 micron regime, is completely hidden from view in the edge-on configuration.
Scattered starlight from the outer disc and/or the envelope accounts for the short wavelength bump of the SED, irrespective of the amount of emission from the disc itself.
Only the quality of the OVRO map at 1.3\,mm provides us with the possibility (and the need) to conclude about the inner clearing.

A possible explanation for a large void with $90\au$ in diameter as we found it here might be, that the disc in the Bok globule CB~26 is actually a circumbinary disc.
Binarity is also a possible explanation for the rotating molecular outflow described by \cite{l08}.
However, a detailed dynamical study is not the scope of this paper.

Of course, another idea might be that the dust in the inner region has already bean processed to planetesimals, or at least to bodies that are large enough to decouple from the disc an its dynamics and do not contribute to the mm-flux.
However, a first indicator for a low age of the system is the fact that it is still deeply embedded in its parental cloud.
Another indication to the young age of the system is the absence of dust grains larger than found in the interstellar medium.
Hence, we do not expect that we may be dealing with a so called ``transitional disc''\footnote{Those discs are considered to be predecessors of evolved debris-type discs.} and a planet population in the centre that has already cleaned up those formerly dusty regions.

\subsection{Disc mass}
Within our model a rather high disc mass is needed to account for the observed flux in the millimetre regime.
With a total of about $3 \times 10^{-3} \msun$ and dust-to-gas ratio of 
\begin{equation}
  \frac{M_{\rm Gas}}{M_{\rm Dust}} \sim 100,
  \label{eqgdrel}
\end{equation}
 we end up with a total disc mass of $\sim 0.3 \msun$ which is close to the star's mass.
Hence, we need to reconsider the mass we computed for the disc.
The derived disc mass essentially builds upon the assumptions we had to made about the dust grain chemistry and shape.

\subsubsection{The dust grain structure, temperature and density}
The radiative transfer in our modelling is performed under the assumption of spherical dust grains to avoid equivocalities.
A possibility to have a lower estimate of the dust mass is to dismiss this assumption.
This very much complicates the radiative transfer calculations since we now need to take into account all the possible fractal grain shapes as well as the spatial orientation of every grain.
So far, we do not have the abilities to do such a simulation.
But we can make a gedankenexperiment on parts of the results.
Of course, one expects strong changes in the scattering behaviour of dust grains.
But besides that, one also can think of a plenitude of dust grains with almost the same absorption cross section as spherical dust grains but with much less mass. \cite{v07} discusses very fluffy particles with a porosity up to $90\%$.

We furthermore want to point out, that the grain density $\rho_{\rm grain}$ is not one of the fitting parameters.
The disc's mass is proportional to this density and the number of grains.
We only can constrain within our study the disc structure, the grain size and their number, but not the density of one grain.
For our investigation of the system we used $\rho_{\rm grain} = 2.5 \,{\rm g}\,{\rm cm}^{-3} $, but it might well be less than this value.
In turn, this will alter our estimate for the disc mass by the same factor.

\subsubsection{The snow line}
For the estimation of the total disc mass, gas and dust, one makes use of the canonical relation (\ref{eqgdrel}).
This relation assumes, that there is some gas and dust of the disc inside the snow line and some outside.
The snow line indicates the largest radius for which the temperature in the disc is high enough to keep water from freezing onto the dust grains.
This line is usually set at a radius where the disc temperature drops to 170 K.
As reported above, the maximum temperature we reach within our disc is about 90 K, and the average is only 16 K.
This means, that our entire disc is outside the snow line.
Thus we need to adjust the dust to gas ratio  (\ref{eqgdrel}) to about 50 as published by \cite{k02}.

However, the very same dust model we made use of in this work was also build upon in other modelling projects \citep[eg.][]{w03a}.
In order to allow comparison of our model for CB~26 with those models, we also give considerations about the disc stability when we keep the dust model.

\subsubsection{Stability, Binarity \& Dynamics}

A criterion for a disc to become unstable was first shown by \cite{t64}.
In order to allow self-gravity of the disc to take over, it requires to satisfy
\begin{equation}
  Q = \frac{c_{\rm s} \kappa}{\pi G \Sigma} \simeq 1
  \label{eqtoomre}
\end{equation}
where $c_{\rm s}$ is the local sound speed, $\kappa$ the angular frequency of the disc, $G$ the gravitational constant and $\Sigma$ the surface density.
For values of $Q$ smaller then unity, the disc is assumed to be gravitationally unstable whereas for $Q$ larger unity the disc is supposed to be stable against gravitational collapse.
\begin{figure}
  \resizebox{\hsize}{!}{
   \includegraphics[width=0.8\textwidth]{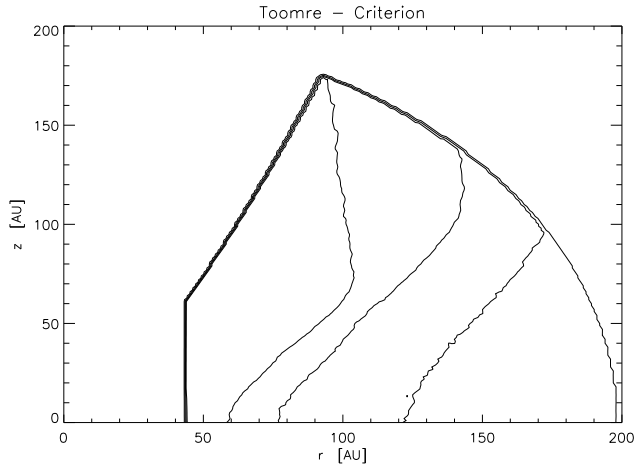}
  }
   \caption{Contour lines of the Toomre parameter $Q$ (see Equation \ref{eqtoomre}). The lines, from right to left, are at levels of $Q= 0.1, \, 0.3, \, 0.5, \,0.7$.}
   \label{pto}
\end{figure}
As Fig. \ref{pto} clearly shows, in our model $Q < 1$ throughout the entire disc by a factor of two to five. 

Yet, the Toomre criterion only can be consistently employed for a system that is formed by a central star and a surrounding disc.
As discussed above, the inner hole we find in CB~26 can be an indication for a binary.
An example for such a system would be the young binary system GG~Tau.
Here, a dusty ring around the central stars has been observed by \cite{g99} with a total ring mass of $0.13\msun$, which is about a factor of two small than the total mass of CB~26.
Yet, having two stars and the disc, the discussion of stability of such a system is quite delicate and not the topic of the present paper.

However, if one accepts to hold on to a single central star, there are still  effects that we did not take care of in our model but might be important for the system's stability.
Following the discussion in the paper of \cite{g01}, discs violating the Toomre criterion are not stable but nevertheless might be in a steady gravoturbulent state.
The paper investigates gravitationally unstable thin Keplerian discs and concludes, that the actual outcome of the instability depends on the cooling  time $\tau_{\rm c}$. 

However,
the parametric ansatz we use does not account for any dynamical interaction within the dust.
Therefore, it would be interesting to couple our radiative transfer code \texttt{MC3D} with hydrodynamic simulations of circumstellar discs
in a future study.
 
\subsection{The width of the dark lane \& envelope structure}
\begin{figure}
  \resizebox{\hsize}{!}{
    \includegraphics[width=0.5\textwidth]{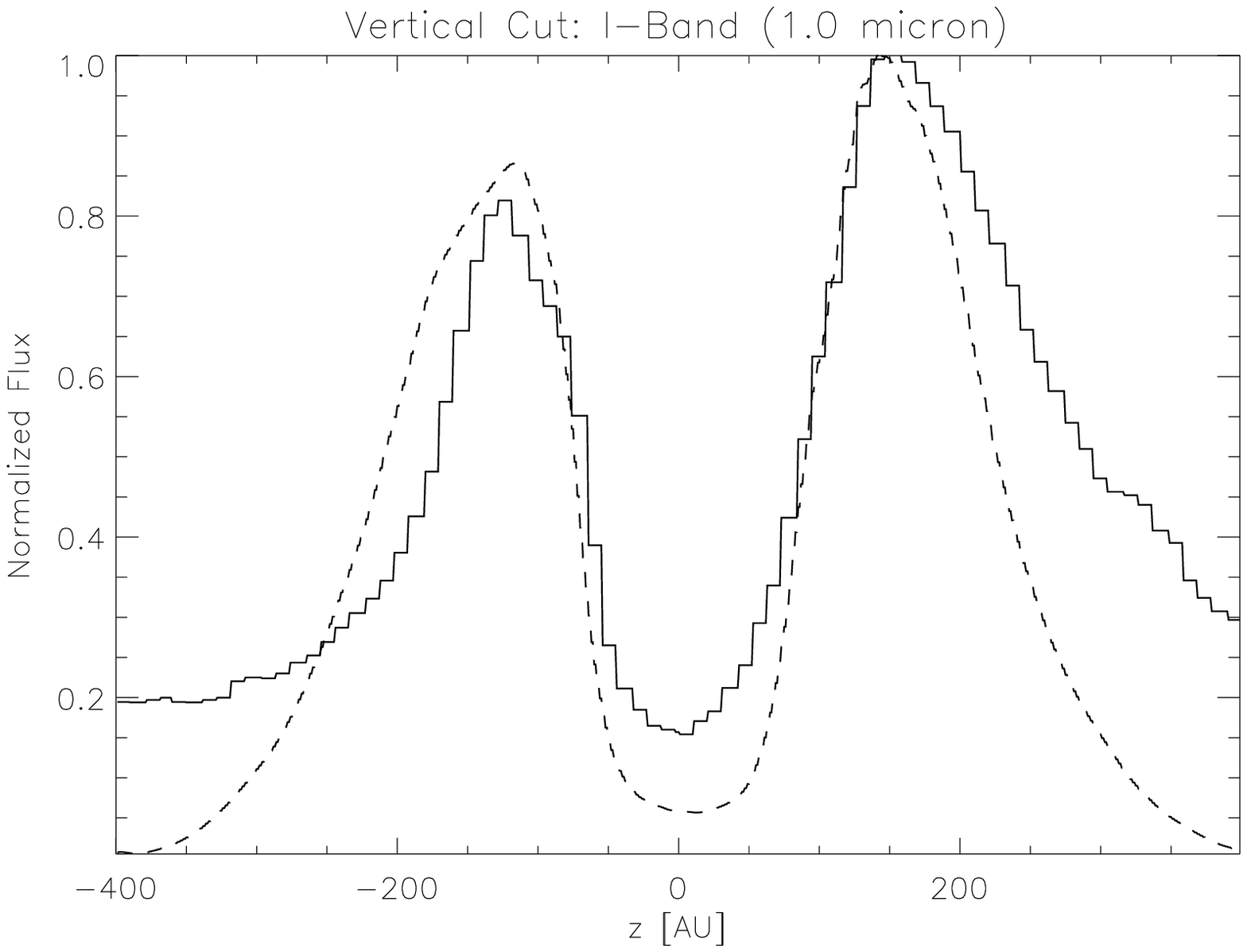}
    \includegraphics[width=0.5\textwidth]{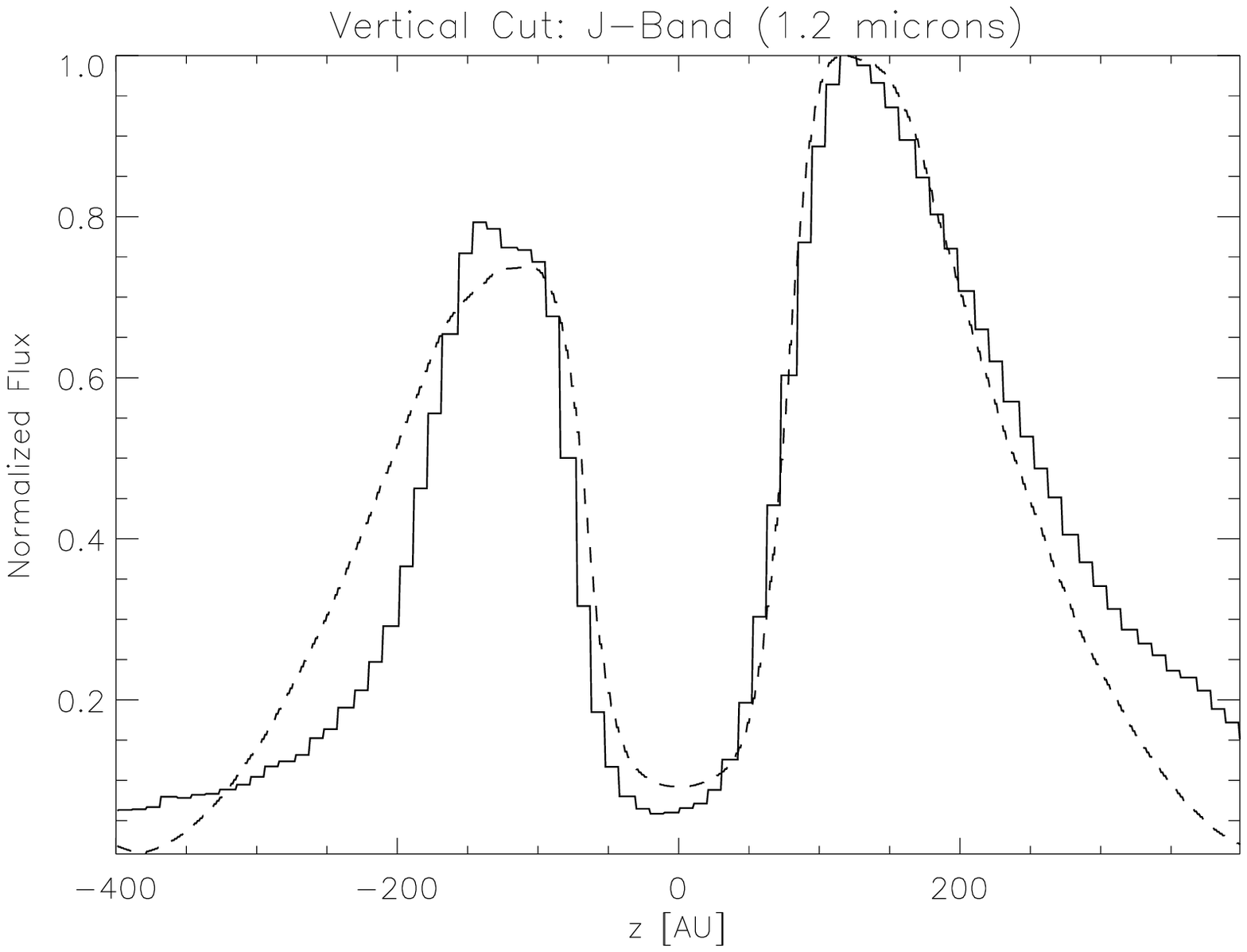}
  }

  \resizebox{\hsize}{!}{
   \includegraphics[width=0.5\textwidth]{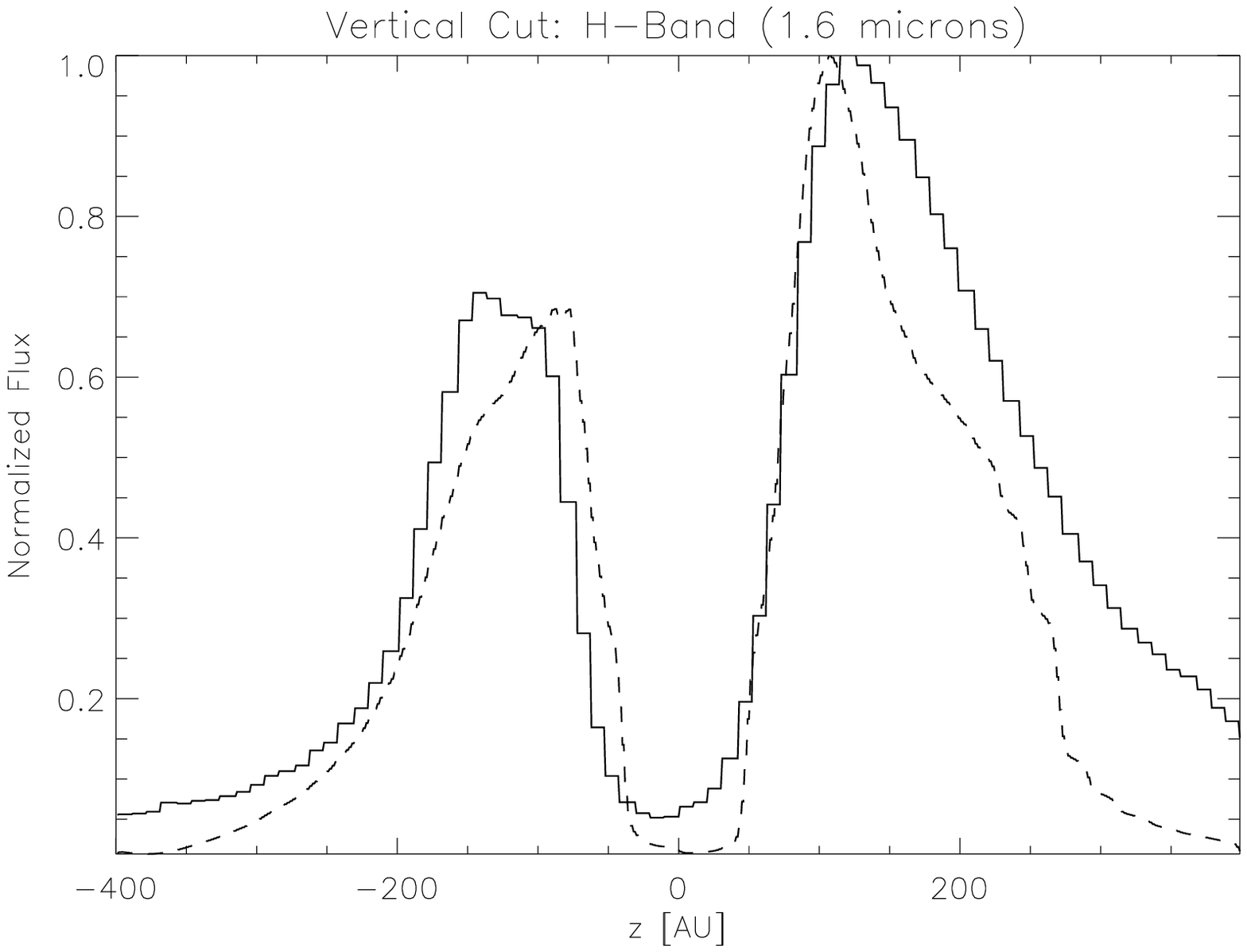} \\
   \includegraphics[width=0.5\textwidth]{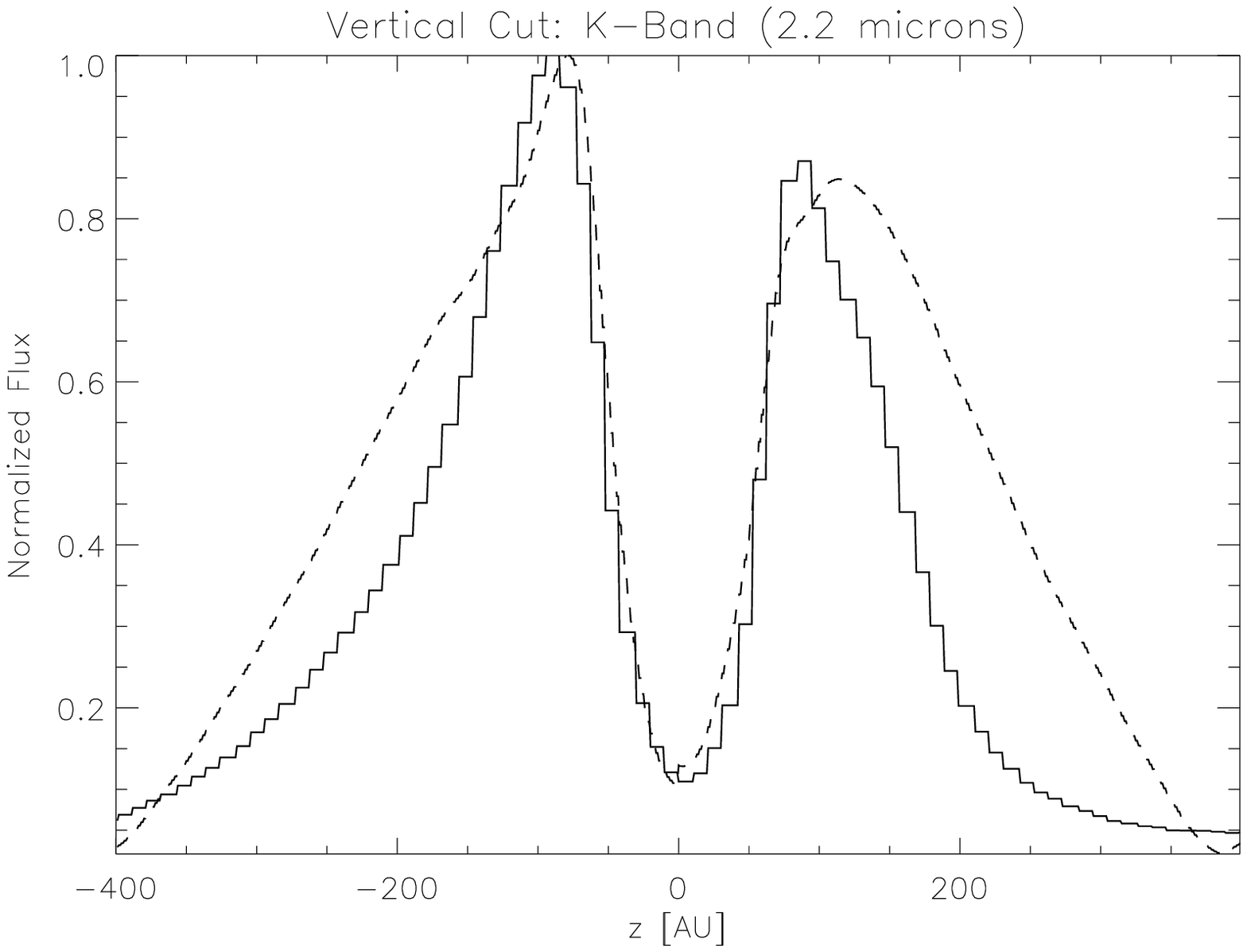} \\
  }
   \caption{
      Vertical cuts at the horizontal centre through the near-infrared scattered light images.
      The solid line is from observational data, the dashed line from our best-fit model.
      For an illustration where those cuts are obtained in the respective maps, see Fig. \ref{hcont}.
      The plots are normalised to 1. In case of the I-Band, this corresponds to $1.3\times10^{-7}\Jy\!/\!\beam$, $5\times10^{-6}\Jy\!/\!\beam$ in the J-Band,  $2\times10^{-5}\Jy\!/\!\beam$ in the H-Band, and  $2.2\times10^{-4}\Jy\!/\!\beam$ (`beam' refers to the FWHM area of the PSF).
   }
   \label{nircuts}
\end{figure}
\begin{figure}
  \resizebox{\hsize}{!}{
   \includegraphics{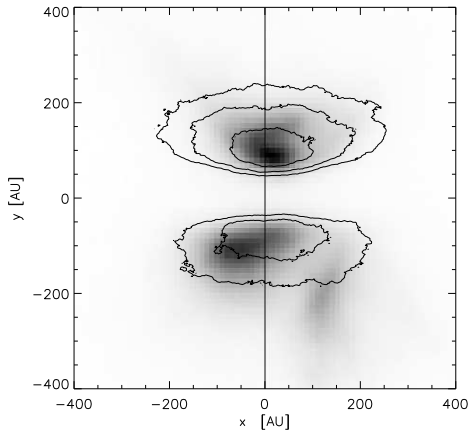}
  }
   \caption{Inverse H-Band image from HST/NICMOS. The contour lines are from our best-fit model. From the outer to the inner lines are they at $27\%$, $41\%$, $68\%$,of the peak height. The second dark lane is not reproduced by our modelling as we did not aim at envelope asymmetries.
The PSF for those images is rather small $\sim 7\au$ FWHM. The vertical line illustrates how the cuts were taken for Fig. \ref{nircuts}.}
   \label{hcont}
\end{figure}

Another aim of our modelling efforts was to mimic the chromaticity of dust lane, which is narrowing with increasing wavelength.
Fig. \ref{nircuts} shows cuts from north to south through the centre of the NIR maps from modelling and observations.
Our model of the spherical envelope is quite successful in reproducing the overall flux at each wavelength.
Also, wavelength-dependent width of the dark lane is correctly reproduced.

Fig. \ref{hcont} shows an overlay of the HST/NICMOS H-band image with the contours given by the best-fit model.
The Fig. clearly shows, that the dark lane is well reproduced.
Since our envelope model, Equation (\ref{eqenvrot}), is axial symmetric, we cannot expect the asymmetries of the lower lobe to be modelled as well.
Thus, the model yields the seen discrepancy to the observation.

A second effect of the aforesaid disregard is that the lower lobe flux, which in the image is concentrated on the left hand side, is in the model distributed among the complete lower lobe. 
Since both fluxes are equal in magnitude we end up with a smaller maximum in the model which also can be seen in Fig. \ref{hcont}.

\subsubsection{Alternative model for the envelope}
\begin{figure}
  \resizebox{\hsize}{!}{
   \includegraphics[width=0.5\textwidth]{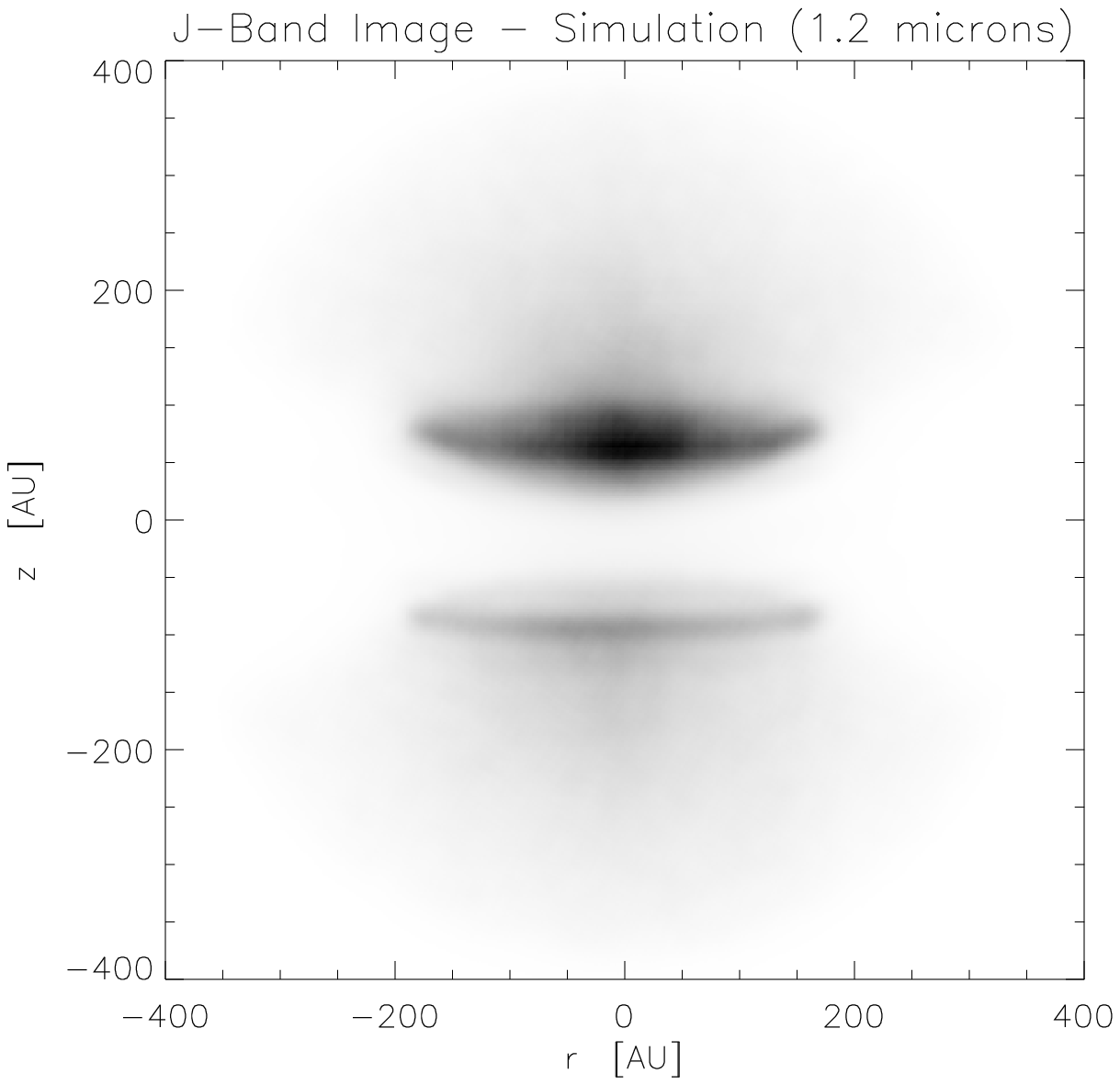}

   \includegraphics[width=0.5\textwidth]{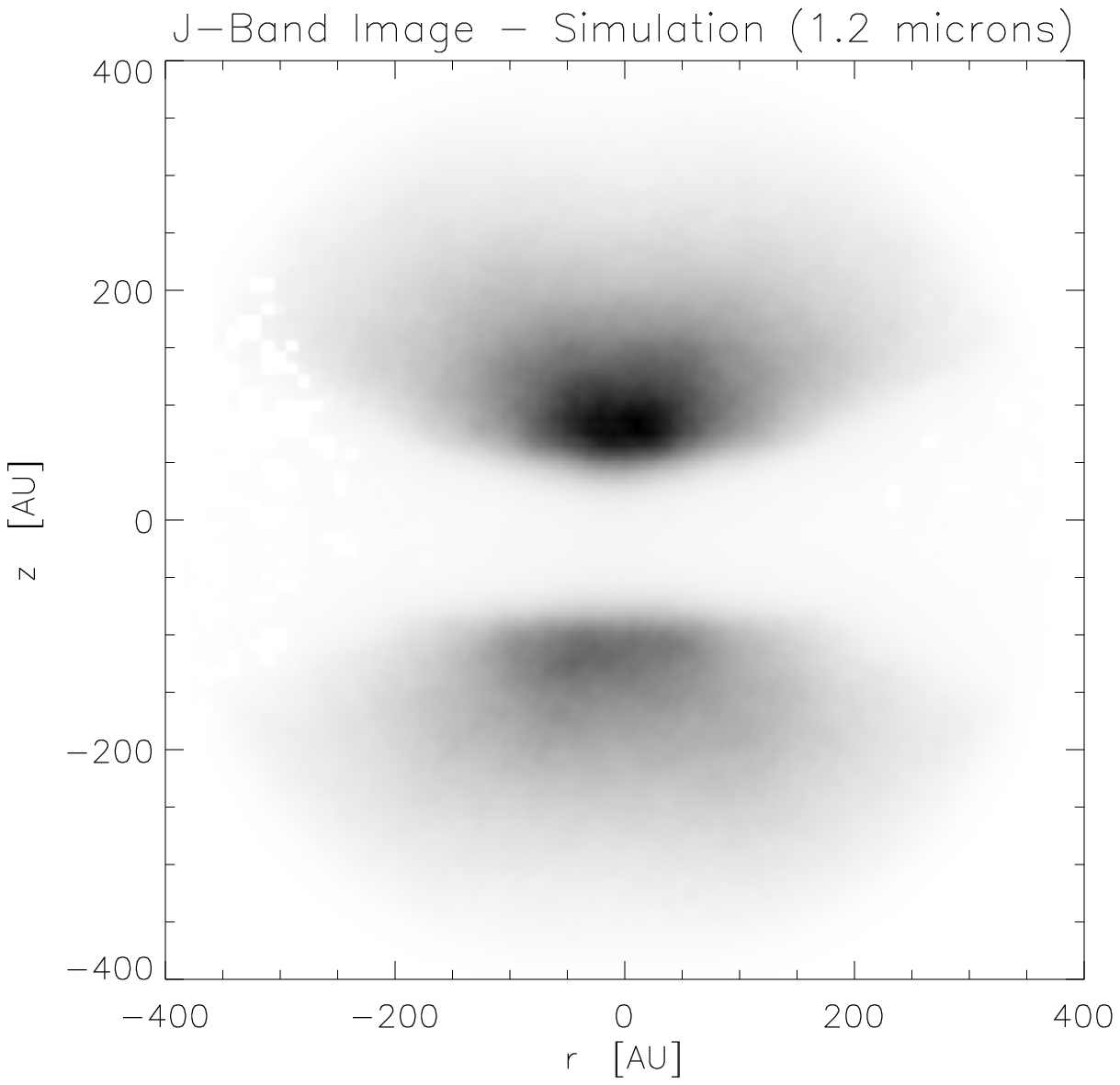}
  }
   \caption{J Band image from a simulation with a power-law envelope structure \emph{(left)} and from rotating envelope \emph{(right)}.
      See also Fig. \ref{NIRmaps} for the observation.
    }
   \label{glowbanana}
\end{figure}

Besides the rotating, in-falling envelope model as described in Equation (\ref{eqenvrot}), we also tested a spherical symmetric density distribution as a model for the envelope:
\begin{equation}
   \rho_\env (\vec{r}) = \rho_{0,\env} \rcyl^\gamma
   \label{eqenvpower}
\end{equation}
This brings two parameters into the model, $\rho_{0,\env}$ and $\gamma$.
With this parameters we are able to model the dependency of the dust lane wideness on wavelength as well as the overall SED.
Yet, this approach has two major drawbacks:
\begin{enumerate}
  \item The spatial flux distribution as seen in the I, J, H, and K images cannot be reproduced. In fact, we obtain a much more concentrated flux distribution above and below the dark lane than what we see in Fig. \ref{HSTradiooverlap}. Fig. \ref{glowbanana} illustrates this.
  \item In the SED appears a strong silicate \emph{emission} feature between $8\mum\le\lambda\le10\mum$. We are not able to have the feature disappearing as required by observations except for the inclination approaching values $75^\circ > \theta$. But this clearly contradicts the edge-on nature of the system.
\end{enumerate}
Therefore, we discarded Equation (\ref{eqenvpower}) as a model for the envelope.

\subsection{Inclination of the disc}
\begin{figure}
  \resizebox{\hsize}{!}{
   \includegraphics[width=0.8\textwidth]{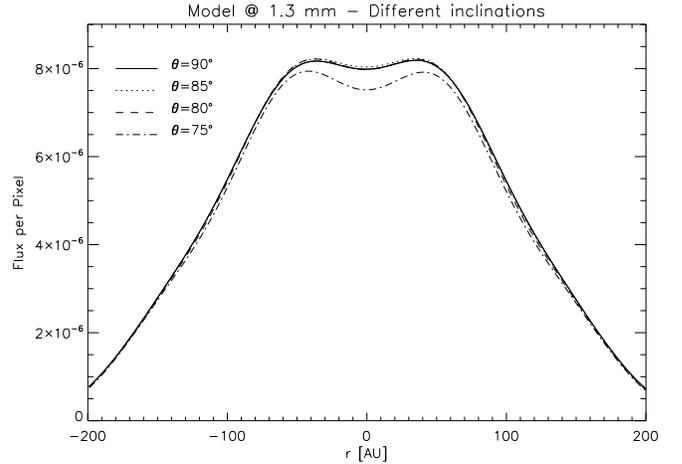}
  }
   \caption{Dependency of the brightness distribution along a central cut on the inclination. All the curves for $\theta = 90^\circ, \, 85^\circ, \, 80^\circ$ overlay. }
   \label{inccut}

\end{figure}
Within our study, we considered two ways to get a hand on the disc's inclination.
The first one is to infer it from the millimetre observations and the second one from the maps in the near-infrared.

In Fig. \ref{inccut} the spatial brightness distribution along a horizontal cut in the 1.3 mm map from the best-fit model at different inclinations is shown.
For inclination values in the range of $\theta = [80^\circ;90^\circ]$ the profile does not change considerably.
For inclinations smaller than $80^\circ$ the profile gets lower.
This is because for an observer the warm inner rim of the disc looks at exactly $90^\circ$ like a line and becomes more and more a circle as one goes from edge-on to face-on orientation.
As long as the complete rim fits into the size of the point spread function, one cannot distinguish between the different inclinations.
But as soon as the rim gets less ellipsoidal and is not longer within the scope of one beam, the flux in the centre decreases.
This is what can be seen in Fig. \ref{inccut}.
Thus, from the comparison of models at different inclinations with the observations, we can infer an value for theta in the aforesaid range of $\theta = [90^\circ;80^\circ]$.
Naturally, this way of reasoning is only valid for optically thin systems.
As outlined earlier, this is the case for CB~26.

The second possibility to conclude about the disc's inclination is to compare the relative peak heights from vertical cuts through the spatial brightness distributions in the near-infrared.
Assuming a symmetric system one expects that both peaks exhibit the same height if the inclination is exactly edge-on.
From an analysis of the actual peak heights (see Fig. \ref{nircuts}) in the HST/NICMOS images we deduce an inclination of $\theta = 85^\circ \pm 5^\circ$ which is in good agreement with the numbers we obtained from the millimetre maps.

\subsection{Comparison with similar objects}
In general our model has disc and envelope parameters that are comparable with those of the circumstellar environment of other young stellar objects.
Yet, there are not many edge-on seen circumstellar discs in the sky for which the same richness of observational data is available for modelling as it is for CB~26.
Hence, the number of comparable multi-wavelength studies on edge-on discs is equally limited.
Two objects, HH~30 and the ``Butterfly-star'', share a large number of features with the disc in CB~26.
All three are seen circumstellar discs, seen almost edge-on.

\cite{w03a} has compiled a similar data set for a model for IRAS 04302+2247, the so called ``Butterfly-star''.
The data set includes millimetre maps and high-resolution near-infrared images obtained with HST/NICMOS. 
Utilising the same techniques as in this study, the authors' main result is that the dust properties must be different in the circumstellar disc and in the envelope.
Whilst a grain size distribution with grain radii up to $100 \mum$ is required to reproduce the millimetre observations of the disc, the envelope is dominated by smaller grains similar to those of the interstellar medium.
This is quite in contrast to our model, where we find grains with almost ISM-like properties are needed in both, disc and envelope.
However, the millimetre maps of the model do not suggest any presence of a large inner void as the spatial resolution is not good enough.
In a later investigation \citep{w08} on the ``Butterfly-star'' with the Sub-Millimetre Array the authors as well discover a ``dip'' in the horizontal brightness distribution.
According to the model however, in this case the effect is to be understood as an effect of the optical depth and not as an inner void.

HH~30 was identified as a circumstellar disc with a large inner void present by HH~30 by \cite{gu8}.
The authors found a inner hole of radius $r_{\rm in}=40 \pm 5 \au$ and an outer disc radius of $r_{\rm out}=128 \pm 3 \au$.
At this numbers HH~30 and CB~26 are quite comparable.
The separation of two central stars is postulated to be about $15\au$, so this might also be good guess for CB~26.
Yet, in contrast to our model for CB~26, no envelope structure is needed to explain the appearance of HH~30 in the near-infrared bands.
This suggests that CB~26 is less dynamically evolved than HH~30.
Both systems drive a molecular outflow, but CB~26 shows clear signatures of outflow rotation \citep{l08}, while HH~30 does not \citep{p06}.
The reason for this difference remains unclear, though the early dynamically state of CB~26 as compared to HH~30 may provide the key.
Yet another difference between HH~30 and our object is that the presence of cm size grains is needed to model the value for $\beta_{\mm}$ \citep{w02}.

As a summary, we see that, although all three objects, CB~26, HH~30, and IRAS~04302+2247 feature the same structure, a detailed investigation as presented in this work shows, that they are systems at different evolutionary states.

\subsection{Errors \& caveats}
We conclude this section by discussing simplifications and resulting possible sources of errors within the framework of our model which have not been mentioned so far.

A matter that has not been touched so far is the uniqueness of or model.
Despite the simplifications we applied the volume of the parameter space is still to vast to be completely scanned in a single study.
We therefore employed the above described step-by-step technique for the parameters to find our best-fit model.

For instance, we started to model the millimetre maps as for those the envelope structure is not a dominant contribution.
Thus, we obtain information about the disc total mass as well as its inner and outer radius and the apparent absence of large dust grains.
The exploration of the near infrared images then showed the requirement for a rotational envelope structure rather then a simple power law distribution.
By treating the parameters in this independent and serial manner, the order in which they are fitted might become an important issue.
For instance, the study could first have been focusing on the near infrared images and wideness of the dust lane.
As we know from our experience from modelling other objects this wideness usually requires small grains at least in the envelope.
Modelling the SED in the millimetre regime as a second step would still have lead us to the overall usage of ISM grains.

This raises the question, if our model is really a global minimum in the parameter space or just a local one and if there exists in that space a model, that reproduces the observations on CB\,26 even better.
We exclude this possibility.
The model we obtained exhibits some unexpected features.
Therefore, in the framework of our model set we thoroughly explored the range of those parameters upon which these features sensitively depend.
For example, we did all the combinations for values of the inner radius and the scale height and disc mass.
If adjust accordingly, all three can produce the plateau observed in the $1.3\mm$ image.
However, only the larger inner radius prevailed.
A small scale height might squeeze the dust tight enough for high optical depth, but clearly contradicts the wavelength dependence of the dark lane in the NIR images.
The same holds for the maximum grain size.
Whereas here we do not even have the possibility to mimic ISM behaviour by means of disc geometry.

As a remaining issue, we need to think about parameters not varied at all in our modelling effort.
As explained in sections above, model assumptions such as spherical grains versus fractal grains can hold the key to the mass problem of the disc.
However, this would not alter the model we have at hand, especially this does not provide a hint to the ``real'' global $\chi^2$-minimum if one thinks the model is trapped in a local minimum.

This might also be an issue for the stellar parameters.
Varying the luminosity and effective temperature of the embedded T~Tauri star in general changes the total energy throughput in the radiative transfer and the location of the peak of the stellar spectrum in the wavelength space.
Deviations from our assumed ``typical'' T~Tauri star of course will affect the numbers of our best-fit model.
For instance, since the total mass critically depends on the flux in the millimetre regime and this flux in turn on the total energy provided by the central star.
Also, the screen introduced to mimic interstellar extinction is affected by the choice of surface temperature.
However, no choice of stellar parameters is able to affect the main conclusions of our model. 
These are the presence of ISM grains in the disc and the inner void.

Despite the discussed caveats, the fact we actually found a good model means that we do not need to call for complex physics, such as grain growth or dust settling.
The data do not require this.

\section{Summary}
For a large span of wavelengths, we have compiled a high quality data set for the circumstellar disc in the Bok globule CB~26.
We obtained images in the near infrared and in the millimetre regime as well as photometric data and spectra.
Together with literature values, we have constructed a detailed model that allows interpretation of observations with one single set of parameters.
The conclusions we obtained are multifarious: 
\begin{enumerate}
  \item In order to account for the brightness distribution in the $1.3 \mm$ map we needed to include an inner hole with a radius of 45 AU.
	Observations to come, especially in the sub-millimetre regime are suitable to confirm our interpretation of a low optical depth along the line of site at $1.3 \mm$.
  \item Our model very nicely reproduces the prominent chromaticity  of the dark lane as seen in the near infrared images.
  \item Based on the chosen dust composition (astronomical silicate, graphite) and the resulting opacity structure of the best-fit disc model, we find that the millimetre SED indicates that the grains feature the same grain size distribution and almost the same upper limit to the grain size as the interstellar medium. 
  \item The disc is massive with a total dust and gas mass of $0.3 \msun$ under the assumption of spherical grains and $\rho_{\rm grain} = 2.5 \,{\rm g}\,{\rm cm}^{-3} $, compared to a mass of the central T~Tauri star with $0.5 \msun$ and therefore possibly, but not inevitably, unstable.
    We discussed that due to possible fractal structure and other effects the real disc mass may be small enough to have a stable system.
\end{enumerate}

\begin{acknowledgements}
The authors thank all members of the GEODE-team for their help in this project.
J. Sauter thanks Owen Matthews, Jens Rodmann, and Arjan Bik for enlightening discussions.
This work is supported by the DFG through the research group 759 ``The Formation of Planets: The Critical First Growth Phase''.
F. Menard thanks financial support from {\sl Programme national de Physique Stellaire} (PNPS) of CNRS/INSU, France and from Agence Nationale pour la Recherche of France under contract ANR-07-BLAN-0221.
This work has been supported by NASA funding from the Space Telescope
Science Institute, HST general observer program 10603; and by NASA
funding from the Jet Propulsion Laboratory, under Spitzer general observer
program 30765.
C. Pinte acknowledges the funding from the European Commission's Seventh Framework Program as a Marie Curie Intra-European Fellow (PIEF-GA-2008-220891).
The Submillimeter Array is a joint project between the Smithsonian Astrophysical Observatory and the Academia Sinica Institute of Astronomy and Astrophysics and is funded by the Smithsonian Institution and the Academia Sinica.
\end{acknowledgements}

\bibliographystyle{aa}
\bibliography{cb26.bib}

\begin{thebibliography}{61}
\expandafter\ifx\csname natexlab\endcsname\relax\def\natexlab#1{#1}\fi

\bibitem[{{Alonso-Albi} {et~al.}(2008){Alonso-Albi}, {Fuente}, {Bachiller},
  {Neri}, {Planesas}, \& {Testi}}]{a08}
{Alonso-Albi}, T., {Fuente}, A., {Bachiller}, R., {et~al.} 2008, \apj, 680,
  1289

\bibitem[{{Beckwith} {et~al.}(1990){Beckwith}, {Sargent}, {Chini}, \&
  {Guesten}}]{B90}
{Beckwith}, S.~V.~W., {Sargent}, A.~I., {Chini}, R.~S., \& {Guesten}, R. 1990,
  \aj, 99, 924

\bibitem[{{Beelen} {et~al.}(2006){Beelen}, {Cox}, {Benford}, {Dowell},
  {Kov{\'a}cs}, {Bertoldi}, {Omont}, \& {Carilli}}]{b06}
{Beelen}, A., {Cox}, P., {Benford}, D.~J., {et~al.} 2006, \apj, 642, 694

\bibitem[{{Bergeron} \& {Dickinson}(2003)}]{b03}
{Bergeron}, L.~E. \& {Dickinson}, M.~E. 2003, {Instrument Science Report NICMOS
  2003-010 (Baltimore: STScI)}

\bibitem[{{Bjorkman} \& {Wood}(2001)}]{b01}
{Bjorkman}, J.~E. \& {Wood}, K. 2001, \apj, 554, 615

\bibitem[{{Bouwman} {et~al.}(2008){Bouwman}, {Henning}, {Hillenbrand}, {Meyer},
  {Pascucci}, {Carpenter}, {Hines}, {Kim}, {Silverstone}, {Hollenbach}, \&
  {Wolf}}]{b08}
{Bouwman}, J., {Henning}, T., {Hillenbrand}, L.~A., {et~al.} 2008, ArXiv
  e-prints, 802

\bibitem[{{Cardelli} {et~al.}(1989){Cardelli}, {Clayton}, \& {Mathis}}]{c89}
{Cardelli}, J.~A., {Clayton}, G.~C., \& {Mathis}, J.~S. 1989, \apj, 345, 245

\bibitem[{{Cashwell} \& {Everett}(1959)}]{c59}
{Cashwell}, E.~D. \& {Everett}, C.~J. 1959, A practical manual on the Monte
  Carlo Method for random walk problems (Pergamon)

\bibitem[{{D'Alessio} {et~al.}(1999){D'Alessio}, {Cant{\'o}}, {Hartmann},
  {Calvet}, \& {Lizano}}]{d99}
{D'Alessio}, P., {Cant{\'o}}, J., {Hartmann}, L., {Calvet}, N., \& {Lizano}, S.
  1999, \apj, 511, 896

\bibitem[{{Dowell} {et~al.}(2003){Dowell}, {Allen}, {Babu}, {Freund},
  {Gardner}, {Groseth}, {Jhabvala}, {Kovacs}, {Lis}, {Moseley}, {Phillips},
  {Silverberg}, {Voellmer}, \& {Yoshida}}]{d03}
{Dowell}, C.~D., {Allen}, C.~A., {Babu}, R.~S., {et~al.} 2003, in Society of
  Photo-Optical Instrumentation Engineers (SPIE) Conference Series, Vol. 4855,
  Society of Photo-Optical Instrumentation Engineers (SPIE) Conference Series,
  ed. T.~G. {Phillips} \& J.~{Zmuidzinas}, 73--87

\bibitem[{{Draine}(2006)}]{d06}
{Draine}, B.~T. 2006, \apj, 636, 1114

\bibitem[{{Draine} \& {Lee}(1984)}]{d84}
{Draine}, B.~T. \& {Lee}, H.~M. 1984, \apj, 285, 89

\bibitem[{{Draine} \& {Lee}(1987)}]{d87}
{Draine}, B.~T. \& {Lee}, H.~M. 1987, \apj, 318, 485

\bibitem[{{Draine} \& {Malhotra}(1993)}]{d93}
{Draine}, B.~T. \& {Malhotra}, S. 1993, \apj, 414, 632

\bibitem[{{Duch{\^e}ne} {et~al.}(2004){Duch{\^e}ne}, {McCabe}, {Ghez}, \&
  {Macintosh}}]{d04}
{Duch{\^e}ne}, G., {McCabe}, C., {Ghez}, A.~M., \& {Macintosh}, B.~A. 2004,
  \apj, 606, 969

\bibitem[{{Eisner} {et~al.}(2005){Eisner}, {Hillenbrand}, {Carpenter}, \&
  {Wolf}}]{e05}
{Eisner}, J.~A., {Hillenbrand}, L.~A., {Carpenter}, J.~M., \& {Wolf}, S. 2005,
  \apj, 635, 396

\bibitem[{{Engelbracht} {et~al.}(2007){Engelbracht}, {Blaylock}, {Su}, {Rho},
  {Rieke}, {Muzerolle}, {Padgett}, {Hines}, {Gordon}, {Fadda},
  {Noriega-Crespo}, {Kelly}, {Latter}, {Hinz}, {Misselt}, {Morrison},
  {Stansberry}, {Shupe}, {Stolovy}, {Wheaton}, {Young}, {Neugebauer},
  {Wachter}, {P{\'e}rez-Gonz{\'a}lez}, {Frayer}, \& {Marleau}}]{e07}
{Engelbracht}, C.~W., {Blaylock}, M., {Su}, K.~Y.~L., {et~al.} 2007, \pasp,
  119, 994

\bibitem[{{Gammie}(2001)}]{g01}
{Gammie}, C.~F. 2001, \apj, 553, 174

\bibitem[{{Glauser} {et~al.}(2008){Glauser}, {M{\'e}nard}, {Pinte},
  {Duch{\^e}ne}, {G{\"u}del}, {Monin}, \& {Padgett}}]{g08}
{Glauser}, A.~M., {M{\'e}nard}, F., {Pinte}, C., {et~al.} 2008, \aap, 485, 531

\bibitem[{{Gordon} {et~al.}(2007){Gordon}, {Engelbracht}, {Fadda},
  {Stansberry}, {Wachter}, {Frayer}, {Rieke}, {Noriega-Crespo}, {Latter},
  {Young}, {Neugebauer}, {Balog}, {Beeman}, {Dole}, {Egami}, {Haller}, {Hines},
  {Kelly}, {Marleau}, {Misselt}, {Morrison}, {P{\'e}rez-Gonz{\'a}lez}, {Rho},
  \& {Wheaton}}]{g07}
{Gordon}, K.~D., {Engelbracht}, C.~W., {Fadda}, D., {et~al.} 2007, \pasp, 119,
  1019

\bibitem[{{Guilloteau} {et~al.}(2008){Guilloteau}, {Dutrey}, {Pety}, \&
  {Gueth}}]{gu8}
{Guilloteau}, S., {Dutrey}, A., {Pety}, J., \& {Gueth}, F. 2008, \aap, 478, L31

\bibitem[{{Guilloteau} {et~al.}(1999){Guilloteau}, {Dutrey}, \& {Simon}}]{g99}
{Guilloteau}, S., {Dutrey}, A., \& {Simon}, M. 1999, \aap, 348, 570

\bibitem[{{Gullbring} {et~al.}(1998){Gullbring}, {Hartmann}, {Briceno}, \&
  {Calvet}}]{g88}
{Gullbring}, E., {Hartmann}, L., {Briceno}, C., \& {Calvet}, N. 1998, \apj,
  492, 323

\bibitem[{{Ho} {et~al.}(2004){Ho}, {Moran}, \& {Lo}}]{h04}
{Ho}, P.~T.~P., {Moran}, J.~M., \& {Lo}, K.~Y. 2004, \apjl, 616, L1

\bibitem[{{Hughes} {et~al.}(2008){Hughes}, {Wilner}, {Kamp}, \&
  {Hogerheijde}}]{h08}
{Hughes}, A.~M., {Wilner}, D.~J., {Kamp}, I., \& {Hogerheijde}, M.~R. 2008,
  \apj, 681, 626

\bibitem[{{Kessler-Silacci} {et~al.}(2006){Kessler-Silacci}, {Augereau},
  {Dullemond}, {Geers}, {Lahuis}, {Evans}, {van Dishoeck}, {Blake}, {Boogert},
  {Brown}, {J{\o}rgensen}, {Knez}, \& {Pontoppidan}}]{k06}
{Kessler-Silacci}, J., {Augereau}, J.-C., {Dullemond}, C.~P., {et~al.} 2006,
  \apj, 639, 275

\bibitem[{{Kokubo} \& {Ida}(2002)}]{k02}
{Kokubo}, E. \& {Ida}, S. 2002, \apj, 581, 666

\bibitem[{{Kov{\'a}cs}(2006)}]{k06b}
{Kov{\'a}cs}, A. 2006, PhD thesis, AA(Caltech), attila@submm.caltech.edu

\bibitem[{{Kov{\'a}cs} {et~al.}(2006){Kov{\'a}cs}, {Chapman}, {Dowell},
  {Blain}, {Ivison}, {Smail}, \& {Phillips}}]{k06c}
{Kov{\'a}cs}, A., {Chapman}, S.~C., {Dowell}, C.~D., {et~al.} 2006, \apj, 650,
  592

\bibitem[{{Launhardt} \& {Henning}(1997)}]{l97}
{Launhardt}, R. \& {Henning}, T. 1997, \aap, 326, 329

\bibitem[{{Launhardt} {et~al.}(2008){Launhardt}, {Pavlyuchenkov}, {Gueth},
  {Chen}, {Dutery}, {Guilloteau}, {Henning}, {Pietu}, {Schreyer}, \&
  {Semenov}}]{l08}
{Launhardt}, R., {Pavlyuchenkov}, Y., {Gueth}, F., {et~al.} 2008, VizieR Online
  Data Catalog, 349, 40147

\bibitem[{{Launhardt} \& {Sargent}(2001)}]{l01}
{Launhardt}, R. \& {Sargent}, A.~I. 2001, \apjl, 562, L173

\bibitem[{{Lucy}(1999)}]{l99}
{Lucy}, L.~B. 1999, \aap, 344, 282

\bibitem[{{Mathis} {et~al.}(1977){Mathis}, {Rumpl}, \& {Nordsieck}}]{mrn}
{Mathis}, J.~S., {Rumpl}, W., \& {Nordsieck}, K.~H. 1977, \apj, 217, 425

\bibitem[{{Natta}(1993)}]{n93}
{Natta}, A. 1993, \apj, 412, 761

\bibitem[{{Ossenkopf} \& {Henning}(1994)}]{o94}
{Ossenkopf}, V. \& {Henning}, T. 1994, \aap, 291, 943

\bibitem[{{Pety} {et~al.}(2006){Pety}, {Gueth}, {Guilloteau}, \&
  {Dutrey}}]{p06}
{Pety}, J., {Gueth}, F., {Guilloteau}, S., \& {Dutrey}, A. 2006, \aap, 458, 841

\bibitem[{{Pinte} {et~al.}(2007){Pinte}, {Fouchet}, {M{\'e}nard}, {Gonzalez},
  \& {Duch{\^e}ne}}]{c07}
{Pinte}, C., {Fouchet}, L., {M{\'e}nard}, F., {Gonzalez}, J.-F., \&
  {Duch{\^e}ne}, G. 2007, \aap, 469, 963

\bibitem[{{Pinte} {et~al.}(2008){Pinte}, {Padgett}, {Menard}, {Stapelfeldt},
  {Schneider}, {Olofsson}, {Panic}, {Augereau}, {Duchene}, {Krist},
  {Pontoppidan}, {Perrin}, {Grady}, {Kessler-Silacci}, {van Dishoeck},
  {Lommen}, {Silverstone}, {Hines}, {Wolf}, {Blake}, {Henning}, \&
  {Stecklum}}]{p08}
{Pinte}, C., {Padgett}, D.~L., {Menard}, F., {et~al.} 2008, ArXiv e-prints, 808

\bibitem[{{Sault} {et~al.}(1995){Sault}, {Teuben}, \& {Wright}}]{s95}
{Sault}, R.~J., {Teuben}, P.~J., \& {Wright}, M.~C.~H. 1995, in Astronomical
  Society of the Pacific Conference Series, Vol.~77, Astronomical Data Analysis
  Software and Systems IV, ed. R.~A. {Shaw}, H.~E. {Payne}, \& J.~J.~E.
  {Hayes}, 433--+

\bibitem[{{Scoville} {et~al.}(1993){Scoville}, {Carlstrom}, {Chandler},
  {Phillips}, {Scott}, {Tilanus}, \& {Wang}}]{s93}
{Scoville}, N.~Z., {Carlstrom}, J.~E., {Chandler}, C.~J., {et~al.} 1993, \pasp,
  105, 1482

\bibitem[{{Shakura} \& {Syunyaev}(1973)}]{s73}
{Shakura}, N.~I. \& {Syunyaev}, R.~A. 1973, \aap, 24, 337

\bibitem[{{Shirley} {et~al.}(2000){Shirley}, {Evans}, {Rawlings}, \&
  {Gregersen}}]{s00}
{Shirley}, Y.~L., {Evans}, II, N.~J., {Rawlings}, J.~M.~C., \& {Gregersen},
  E.~M. 2000, \apjs, 131, 249

\bibitem[{{Stansberry} {et~al.}(2007){Stansberry}, {Gordon}, {Bhattacharya},
  {Engelbracht}, {Rieke}, {Marleau}, {Fadda}, {Frayer}, {Noriega-Crespo},
  {Wachter}, {Young}, {M{\"u}ller}, {Kelly}, {Blaylock}, {Henderson},
  {Neugebauer}, {Beeman}, \& {Haller}}]{s07}
{Stansberry}, J.~A., {Gordon}, K.~D., {Bhattacharya}, B., {et~al.} 2007, \pasp,
  119, 1038

\bibitem[{{Stapelfeldt} {et~al.}(1998){Stapelfeldt}, {Krist}, {Menard},
  {Bouvier}, {Padgett}, \& {Burrows}}]{s98}
{Stapelfeldt}, K.~R., {Krist}, J.~E., {Menard}, F., {et~al.} 1998, \apjl, 502,
  L65+

\bibitem[{{Stapelfeldt} {et~al.}(2003){Stapelfeldt}, {M{\'e}nard}, {Watson},
  {Krist}, {Dougados}, {Padgett}, \& {Brandner}}]{s03}
{Stapelfeldt}, K.~R., {M{\'e}nard}, F., {Watson}, A.~M., {et~al.} 2003, \apj,
  589, 410

\bibitem[{{Stecklum} {et~al.}(2004){Stecklum}, {Launhardt}, {Fischer},
  {Henden}, {Leinert}, \& {Meusinger}}]{s04}
{Stecklum}, B., {Launhardt}, R., {Fischer}, O., {et~al.} 2004, \apj, 617, 418

\bibitem[{{Toomre}(1964)}]{t64}
{Toomre}, A. 1964, \apj, 139, 1217

\bibitem[{{Ulrich}(1976)}]{u76}
{Ulrich}, R.~K. 1976, \apj, 210, 377

\bibitem[{{Voshchinnikov}(2002)}]{v02}
{Voshchinnikov}, N.~V. 2002, in Optics of Cosmic Dust, ed. G.~{Videen} \&
  M.~{Kocifaj}, 1

\bibitem[{{Voshchinnikov} {et~al.}(2007){Voshchinnikov}, {Videen}, \&
  {Henning}}]{v07}
{Voshchinnikov}, N.~V., {Videen}, G., \& {Henning}, T. 2007, \ao, 46, 4065

\bibitem[{{Watson} \& {Stapelfeldt}(2004)}]{w04}
{Watson}, A.~M. \& {Stapelfeldt}, K.~R. 2004, \apj, 602, 860

\bibitem[{{Weingartner} \& {Draine}(2001)}]{w01}
{Weingartner}, J.~C. \& {Draine}, B.~T. 2001, \apj, 548, 296

\bibitem[{{Whitney} {et~al.}(2003){Whitney}, {Wood}, {Bjorkman}, \&
  {Wolff}}]{w03b}
{Whitney}, B.~A., {Wood}, K., {Bjorkman}, J.~E., \& {Wolff}, M.~J. 2003, \apj,
  591, 1049

\bibitem[{{Wolf}(2003{\natexlab{a}})}]{w03d}
{Wolf}, S. 2003{\natexlab{a}}, \apj, 582, 859

\bibitem[{{Wolf}(2003{\natexlab{b}})}]{w03c}
{Wolf}, S. 2003{\natexlab{b}}, Computer Physics Communications, 150, 99

\bibitem[{{Wolf} {et~al.}(1999){Wolf}, {Henning}, \& {Stecklum}}]{w99}
{Wolf}, S., {Henning}, T., \& {Stecklum}, B. 1999, \aap, 349, 839

\bibitem[{{Wolf} {et~al.}(2003){Wolf}, {Padgett}, \& {Stapelfeldt}}]{w03a}
{Wolf}, S., {Padgett}, D.~L., \& {Stapelfeldt}, K.~R. 2003, \apj, 588, 373

\bibitem[{{Wolf} {et~al.}(2008){Wolf}, {Schegerer}, {Beuther}, {Padgett}, \&
  {Stapelfeldt}}]{w08}
{Wolf}, S., {Schegerer}, A., {Beuther}, H., {Padgett}, D.~L., \& {Stapelfeldt},
  K.~R. 2008, \apjl, 674, L101

\bibitem[{{Wood} {et~al.}(2002){Wood}, {Wolff}, {Bjorkman}, \& {Whitney}}]{w02}
{Wood}, K., {Wolff}, M.~J., {Bjorkman}, J.~E., \& {Whitney}, B. 2002, \apj,
  564, 887

\bibitem[{{Wu} {et~al.}(2007){Wu}, {Dunham}, {Evans}, {Bourke}, \&
  {Young}}]{w07}
{Wu}, J., {Dunham}, M.~M., {Evans}, II, N.~J., {Bourke}, T.~L., \& {Young},
  C.~H. 2007, \aj, 133, 1560

\end{thebibliography}

\end{document}